\documentclass[referee]{aa} 

\usepackage{graphicx}
\usepackage{txfonts}
\usepackage{supertabular}
\begin{document}
   \title{2D metallicity distribution of the ionized gas in NGC~628 and NGC~6946}
   \authorrunning {Cedr\'es et al.}

 \author{Bernab\'e Cedr\'es
        \inst{1,2}
        \and
        Jordi Cepa
        \inst{1,2}
        \and
        \'Angel Bongiovanni
        \inst{1,2}
        \and
        H\'ector Casta\~neda
        \inst{3}
        \and
        Miguel S\'anchez-Portal
        \inst{4}
        \and
        Akihiko Tomita
        \inst{5}}
  \institute{Instituto de Astrof\'{\i}sica de Canarias (IAC), E-38200 La Laguna, Tenerife, Spain\\
  \email{bce@iac.es}
   \and
   Departamento de Astrof\'{\i}sica, Universidad de La Laguna (ULL), E-38205 La Laguna, Tenerife, Spain
   \and
   Departamento de F\'{\i}sica, Escuela Superior de F\'sica y Matem\'atica, IPN, M\'exico D.F., M\'exico
   \and
   Herschel Science Centre, INSA/ESAC, Madrid, Spain
   \and
   Faculty of Education, Wakayama University, Wakayama 640-8510, Japan}
 
\date{Received; accepted}

  \abstract
   {}
   {We present here two H~II region catalogues with azimuthal resolution for the two grand design galaxies NGC~628 and NGC~6946. With the help of these catalogues, we pretend to study several of the properties of the star forming processes that happen in the spiral galaxies.}
   {We have obtained direct imaging in the narrow-band filters centered at H$\alpha$, H$\beta$, [O~II]$\lambda$3727, [O~III]$\lambda\lambda$4959,5007 and their respective continua. After the calibration and correction of the data, we obtain the dereddened fluxes in the aforementioned lines, the sizes, the H$\alpha$ equivalent width, and using two different empirical calibrations, the metallicity. Employing a method based on the Delaunay triangulation, a 2D representation of the metallicity was obtained.}
   {Data for 209 H~II regions of NGC~628 and 226 H~II regions of NGC~6946 were obtained. The radial behaviours of the H$\alpha$ equivalent width, the excitation and, the oxygen abundance were derived. A two dimensional representation of the metallicity and the parameter $P$ was calculated for the galaxies in the study. A comparison between the two empirical calibrations for the metallicity was done.}
   {The behaviour of the extinction and the H$\alpha$ equivalent width is similar to the ones presented in the literature. The oxygen abundance gradients obtained in this study coincide with the previous ones published. However, more regions than in previous studies were obtained. A difference about 0.6\,dex between the two empirical calibrations employed has been found. Finally, the 2D representations of the metallicity show for NGC~628 high metallicity knots, and for NGC~6946 a high metallicity azimuthal structure has been discovered. This high metallicity regions seem to be linked to the arms of the galaxies, and are probably produced by an increase of the temperature of the ionizing clusters in the H~II regions, that may be linked to variations in the IMFs between the arm and interarm regions.}


   \maketitle
%

\section{Introduction}
The H~II regions in grand design galaxies are excellent probes of the physical processes that drive star formation in the arms of those galaxies. To this aim, it is mandatory to determine the parameters that govern the H~II condition. 
Since the pioneering works of Searle (1971) and Hodge (1976), the H~II regions have been used to estimate the oxygen abundance in galaxies (Smith, 1975; Rosa, 1981; McCall et al., 1985; Kennicutt et al., 2003; or Moustakas et al., 2010 among others), to calculate the star formation rate (SFR) and the star formation history, as proposed by Kennicutt (1998), or even variations in the initial mass function (IMF) associated to the presence of density waves (Cedr\'es et al. 2005).

Indeed, the studies with data from H~II regions in spiral galaxies in several wavelengths have proven useful in order to develop further studies (for example Hodge, 1976;  McCall et al., 1985; Belley \& Roy, 1992; Cedr\'es \& Cepa, 2002; Rozas, 2008; Moustakas et al., 2010). Moreover, imaging studies give a 2D information that it is not present in long-slit works, and that may be important in order to understand the physical processes that take place in the galaxy (Cedr\'es et al., 2005; Rosales-Ortega et al., 2011; Cedr\'es et al. 2012).  

The grand design galaxy NGC~628, with arm class 9 from Elmegreen \& Elmegreen (1987) classification, has a low inclination (6.5 degrees, according to Fathi et al. 2007) and presents a large number of bright H~II regions (Table \ref{galaxias}). It has been a main target for the study of H~II regions since the early works of Hodge (1976) and Kennicutt \& Hodge (1976, 1980). It was first reported to have an abundance gradient by Talent (1983). Other spectroscopic works which include this galaxy are: McCall et al. (1985), Ferguson et al. (1998), Bresolin et al. (1999), and Castellanos et al. (2002), among others.
The first narrow band study in this galaxy was done by Belley \& Roy (1992) (thereafter B\&R92), which obtained data up to 130 H~II regions in several emission lines, including H$\alpha$, H$\beta$, [OIII] and [NII]. Recently, a more detailed study in the number of emission lines, using integral field spectroscopy, has been carried out in S\'anchez et al. (2011) and Rosales-Ortega et al (2011) (thereafter RO11) with data for 108 H~II regions. They were also the firsts to do a proper 2D spectroscopic study of the oxygen abundance in a spiral galaxy.

The galaxy NGC~6946 is also a grand design spiral, with the same arm class as NGC~628. However, in optical wavelenghts and in K-band (Regan \& Vogel, 1995), NGC~6946 presents a more complex shape with multiple arms, which leads to suggest that it is more flocculent than NGC~628 (Foyle et al., 2010). It has a larger inclination (30 degrees from Sofue et al., 1999) than NGC~628, and also presents a larger number of bright H~II regions (Table \ref{galaxias}).
This galaxy has been also a primary target for cataloging H~II regions in spiral galaxies: McCall et al. (1985), Ferguson et al. (1998), B\&R92, Moustakas et al. (2010) among others. Unfortunately, for this galaxy there are less works with high spectroscopical resolution, with Garc\'{\i}a-Benito et al. (2010) being the most destacable, but unfortunately in only covers a few H~II regions in the outer parts of the galaxy.

In this work we present the data collected using narrow-band imaging techniques for NGC~628 and NGC~6946, in the emission lines H$\alpha$, H$\beta$, [O~II]$\lambda$3727\,\AA, [O~III]$\lambda$5007\,\AA\ and their respective continua. In section 2 we describe the data obtention, the reduction and the correction processes. In section 3 we present the resultant H~II catalogue and we explore the results for several parameters, such as the H$\alpha$ equivalent width, the extinction and the sizes of the H~II regions. In section 4 we deal with the problem of the determination of the oxygen abundance for both galaxies, and we present a 2D model for the metallicity. In section 5 a summary is presented.

\section{Data}
The data were obtained in a run on 24th to 26th September 2005 at the 2.5\,m NOT, located at Roque de los Muchachos Observatory. The instrument used was ALFOSC in direct imaging mode with narrowband filters centered at the lines [O~II]$\lambda$3727\,\AA, [O~III]$\lambda$5007\,\AA, H$\alpha$ and H$\beta$ and their respective continua. The filters used are described in Table \ref{filtros}. The spatial scale given by NOT with ALFOSC and E2V CCD (2048 $\times$ 2048 pixels) was 0.19$^{\prime\prime}$/pixel. The seeing was between 0.9$^{\prime\prime}$ and 1$^{\prime\prime}$. The integration times for both galaxies in each filter are indicated in Table \ref{tiempos}.

\begin{table*}
\caption{Parameters of the galaxy sample}
\begin{tabular}{c c c c c c c}
\hline\hline
Galaxy & Morphological class & D (Mpc) & R$_{25}$ (arcmin) & V (km/s) & P.A. (degrees) & i (degrees)\\
\hline
NGC~628 & SA(s)c (1) & 7.3 (2) & 5.23 (1) & 657 (3) & 11.8 (4) & 6.5 (5)\\
NGC~6946 & SAB(rs)cd (1) & 5.5 (2) & 5.74 (1) & 40 (6) & 64 (7) & 30 (7)\\
\hline\hline
\end{tabular}
\tablefoot{(1) de Vaucouleurs et al. (1991); (2) Helfer et al. (2003); (3) Lu et al. (1993); (4) Egusa et al. (2009); (5) Fathi et al. (2007); (6) Epinat et al (2008); (7) Sofue et al. (1999).}
\label{galaxias}
\end{table*}

\begin{table*}
\caption{Employed filters}
\begin{tabular}{c c c c}
\hline\hline
Line & $\lambda_c$ (\AA) & FWHM (\AA) & Maximun transmitance (\%)\\
\hline
[O~II] continuum & 3578 & 25 & 48 \\

[O~II] & 3728 & 32 & 27\\
H$\beta$ continuum (Str b) & 4670 & 180 & 89\\
H$\beta$ & 4873 & 45 & 58\\

[O~III] & 5010 & 43 & 56\\

[O~III] continuum & 5105 & 37 & 55\\
H$\alpha$ & 6577 & 180 & 77\\
H$\alpha$ continuum & 6788 & 45 & 73\\
\hline\hline
\end{tabular}
\label{filtros}
\end{table*}

\begin{table*}
\caption{Total integration times at each line and continuum}
\begin{tabular}{c c c c c c c c c}
\hline\hline
 & H$\alpha$ & H$\alpha_{c}$ & H$\beta$ & H$\beta_{c}$ & [OII] & [OII]$_c$ & [OIII] & [OIII]$_c$ \\
\hline
NGC~628 & 3x600\,s & 5x600\,s & 3x1200\,s & 3x600\,s & 4x900\,s & 4x1200\,s & 3x900\,s & 3x900\,s \\
NGC~6946 & 3x600\,s & 3x600\,s & 3x1200\,s & 3x600\,s & 3x900\,s & 3x1200\,s & 3x900\,s & 3x900\,s \\
\hline\hline
\end{tabular}
\label{tiempos}
\end{table*}

\subsection{Reduction and calibration}

The reduction and calibration of the data were carried out using the IRAF\footnote{IRAF is distributed by the National Optical Astronomy Observatory, which is operated by the Association of Universities for Research in Astronomy (AURA) under cooperative agreement with the National Science Foundation.} package. The images were bias substracted employing bias frames taken during the night and the overscan zone of each image. The images were then flatfield corrected using sky flatfields (3 to 4 for each filter), taken during twilight at the beginning and at the end of the observing nights. To correct the images from the sky emission, we obtained the mean value in 20 $\times$ 20 pixels boxes in zones not affected by the galaxy, bright stars or cosmic rays.

The images were calibrated employing spectrophotometric standard stars from Oke \& Gunn (1983) and Oke (1990), and following the procedure described in Barth et al. (1994).

After the calibration, we matched and combined all the images of the galaxy in the same filter using the median. In this way, we were able to wipe out the cosmic rays and to increase the signal to noise of the final images.

\begin{figure*}
 \centering
 \begin{tabular}{c c}
 \includegraphics[width=8cm]{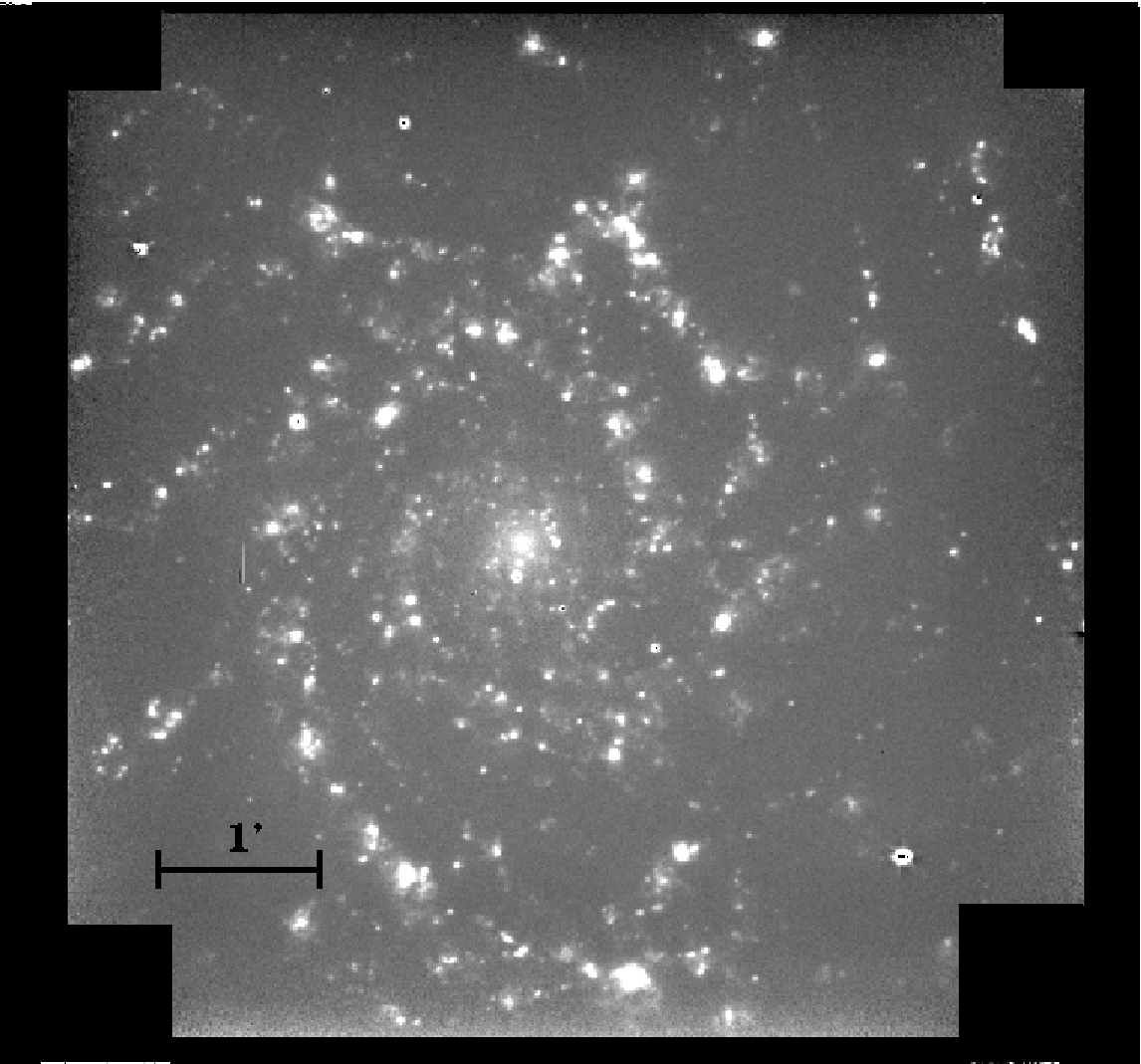} & \includegraphics[width=8cm]{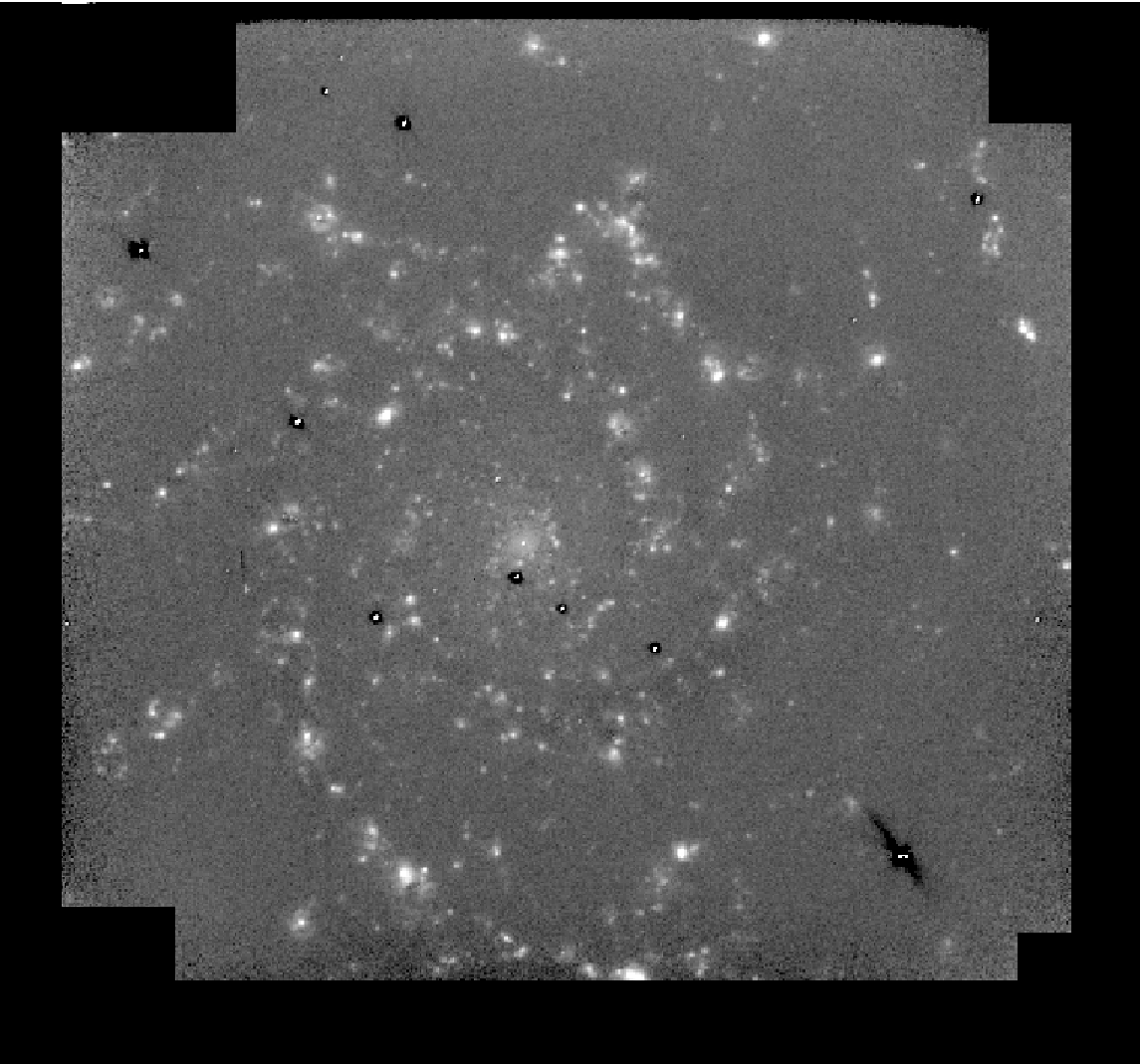} \\
 \includegraphics[width=8cm]{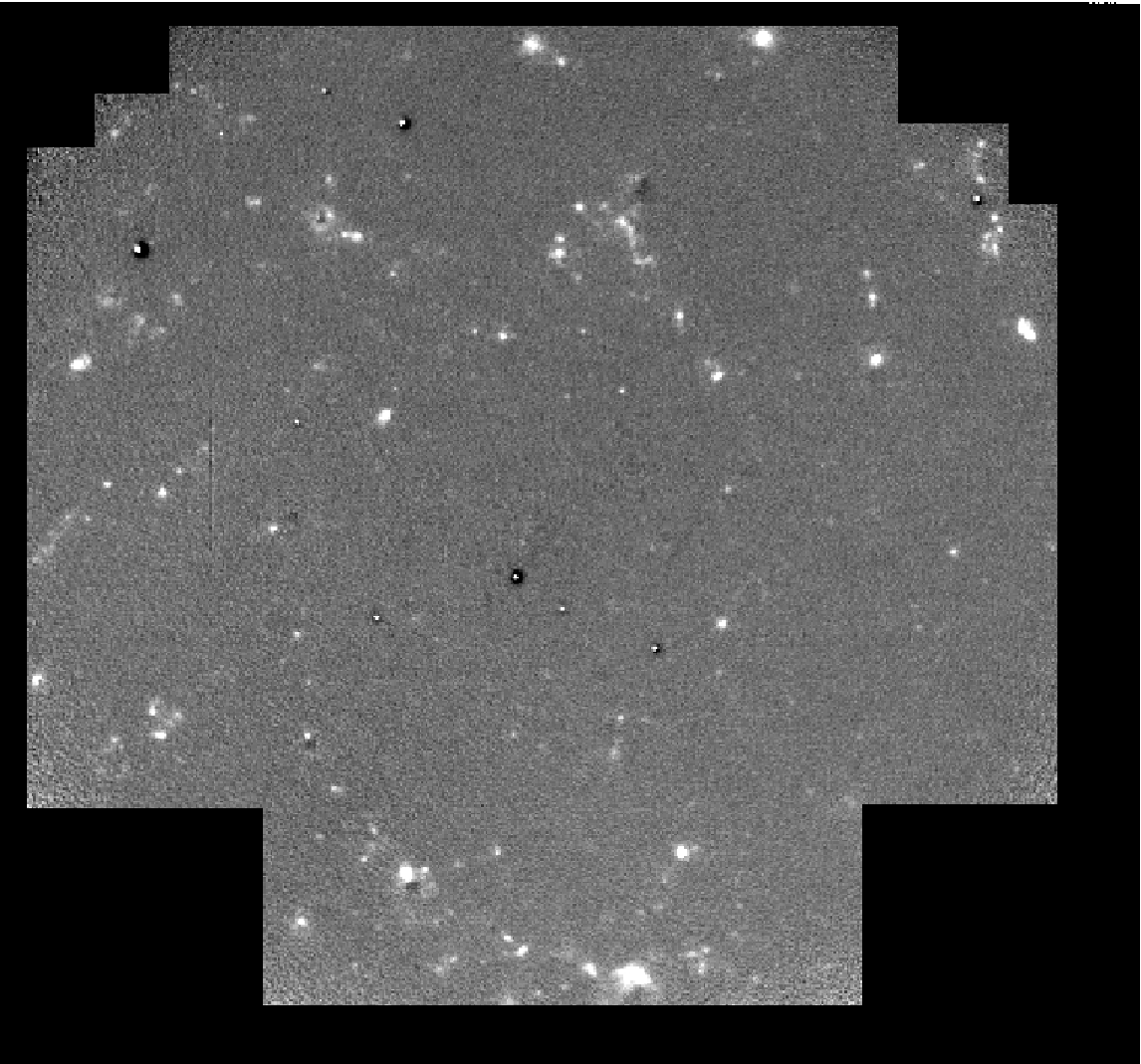} & \includegraphics[width=8cm]{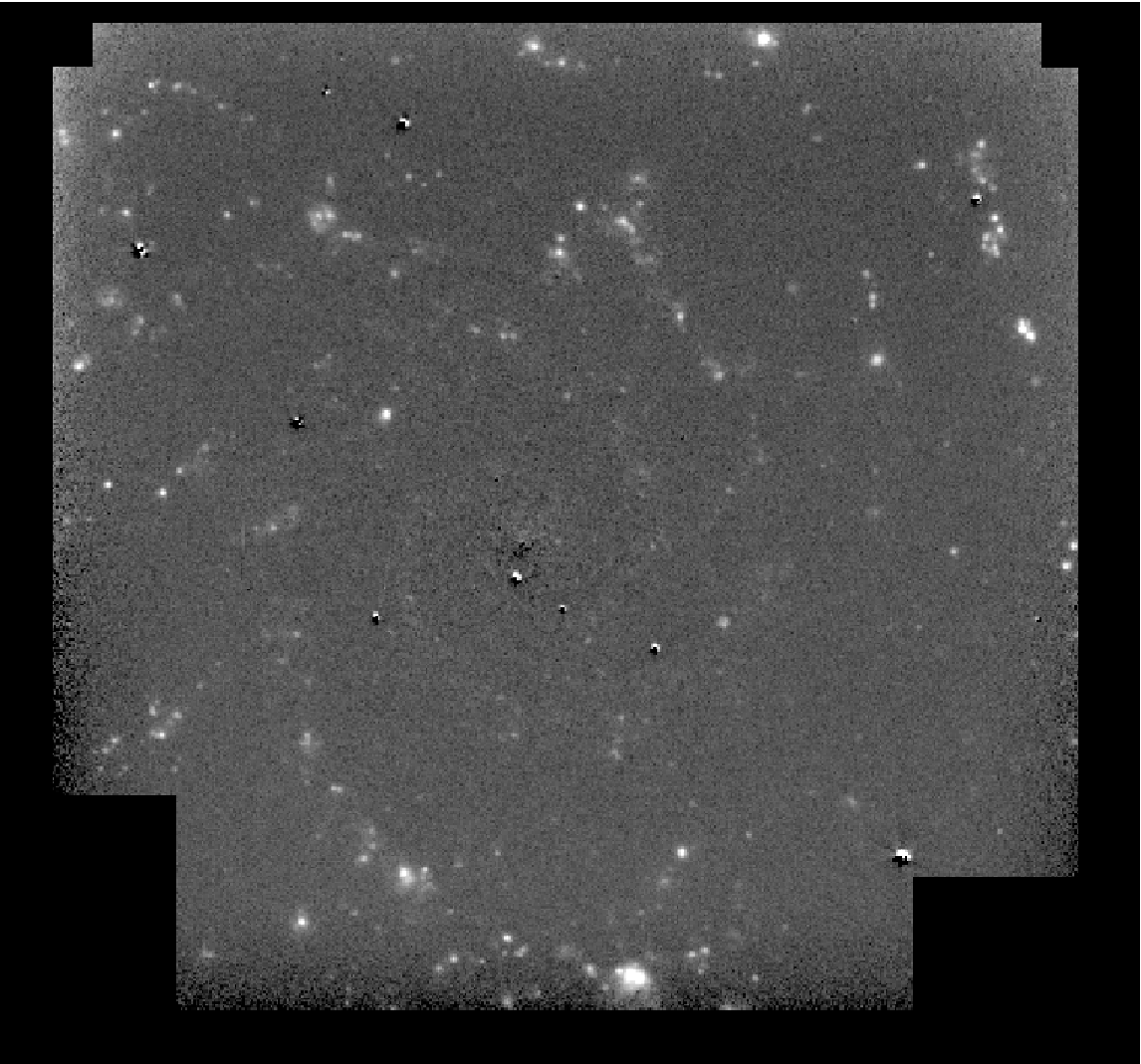} \\
 \end{tabular}
 \caption{Continuum substracted images for NGC~628. Top left, H$\alpha$; top right, H$\beta$; botton left, [OII]; and bottom right [OIII]. North is at top and East at the left for all images. Scale as indicated.}
 \label{im62}
\end{figure*}

\begin{figure*}
 \centering
 \begin{tabular}{c c}
 \includegraphics[width=8cm]{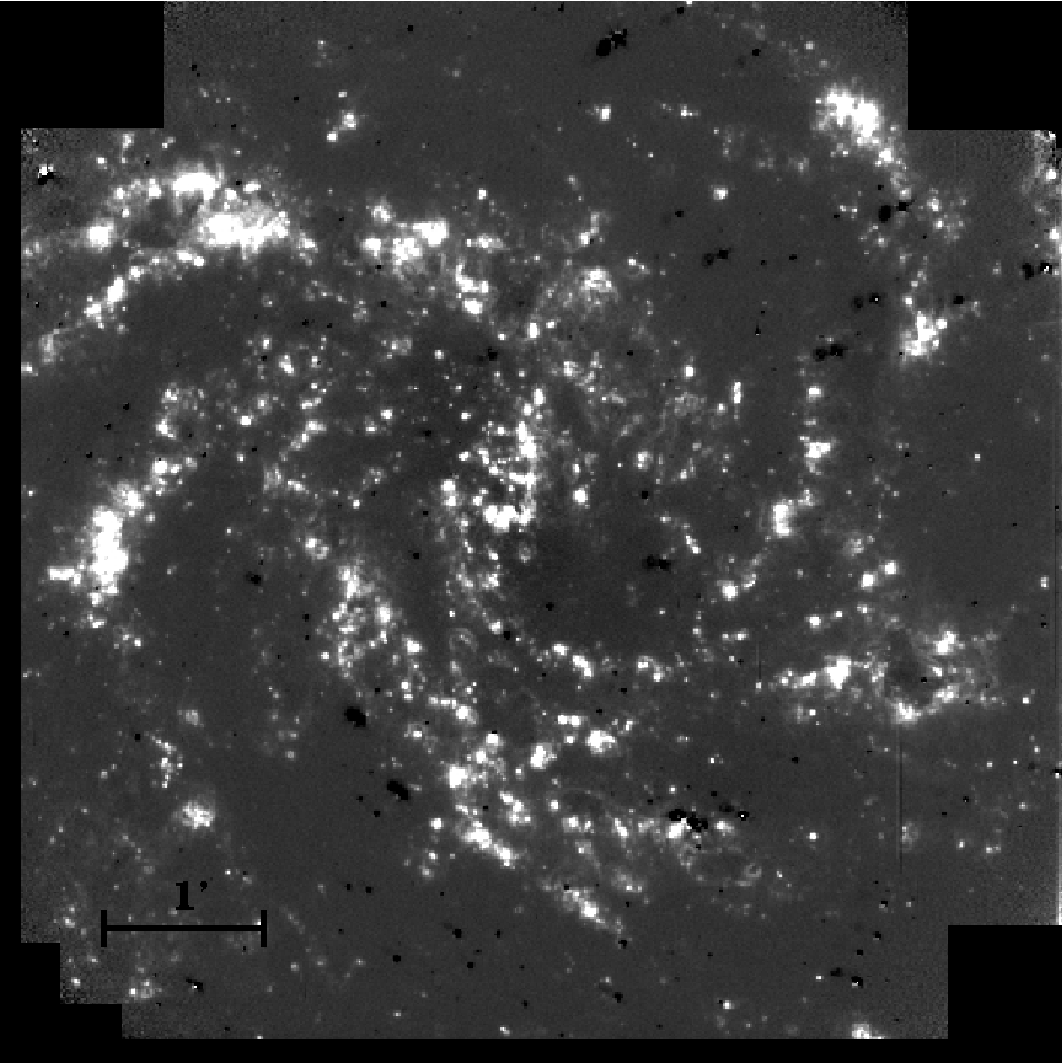} & \includegraphics[width=8cm]{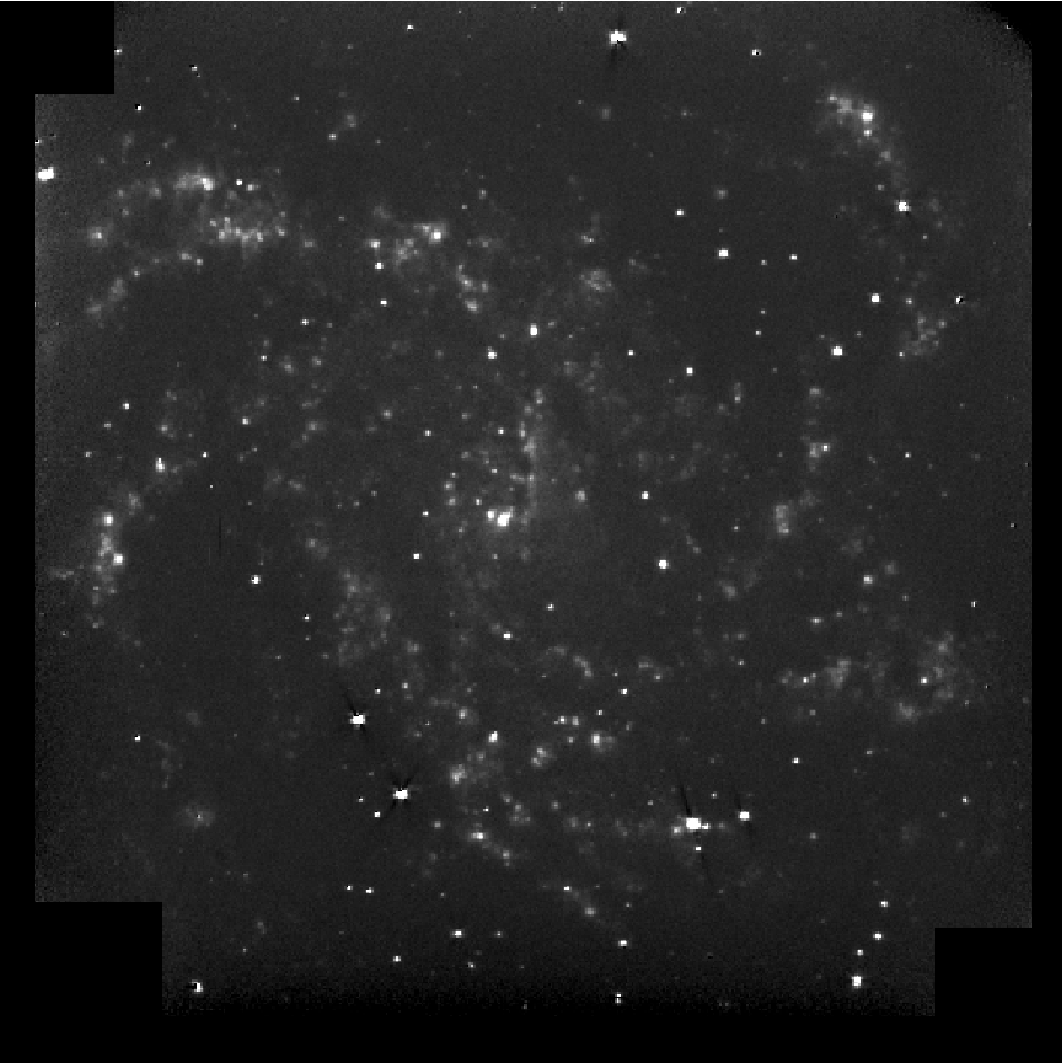} \\
 \includegraphics[width=8cm]{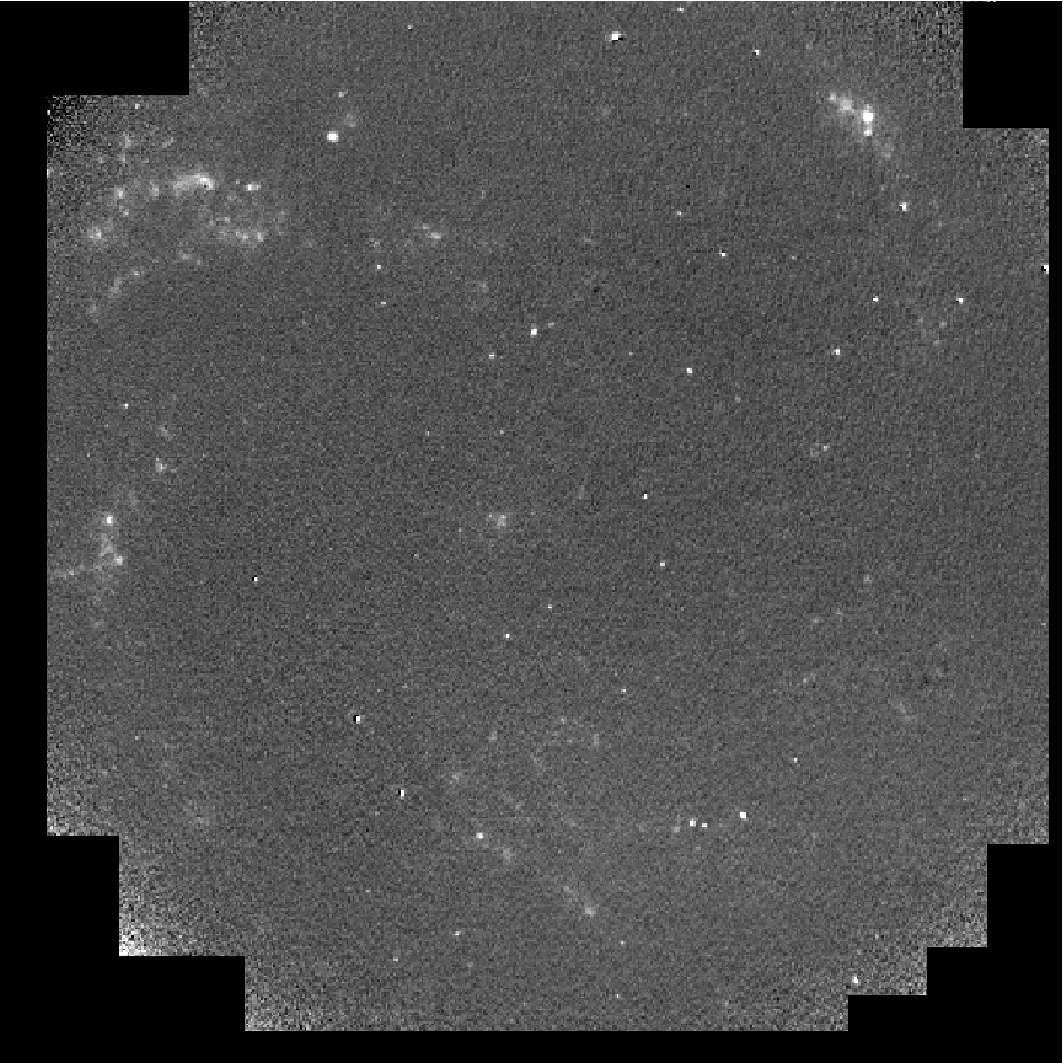} & \includegraphics[width=8cm]{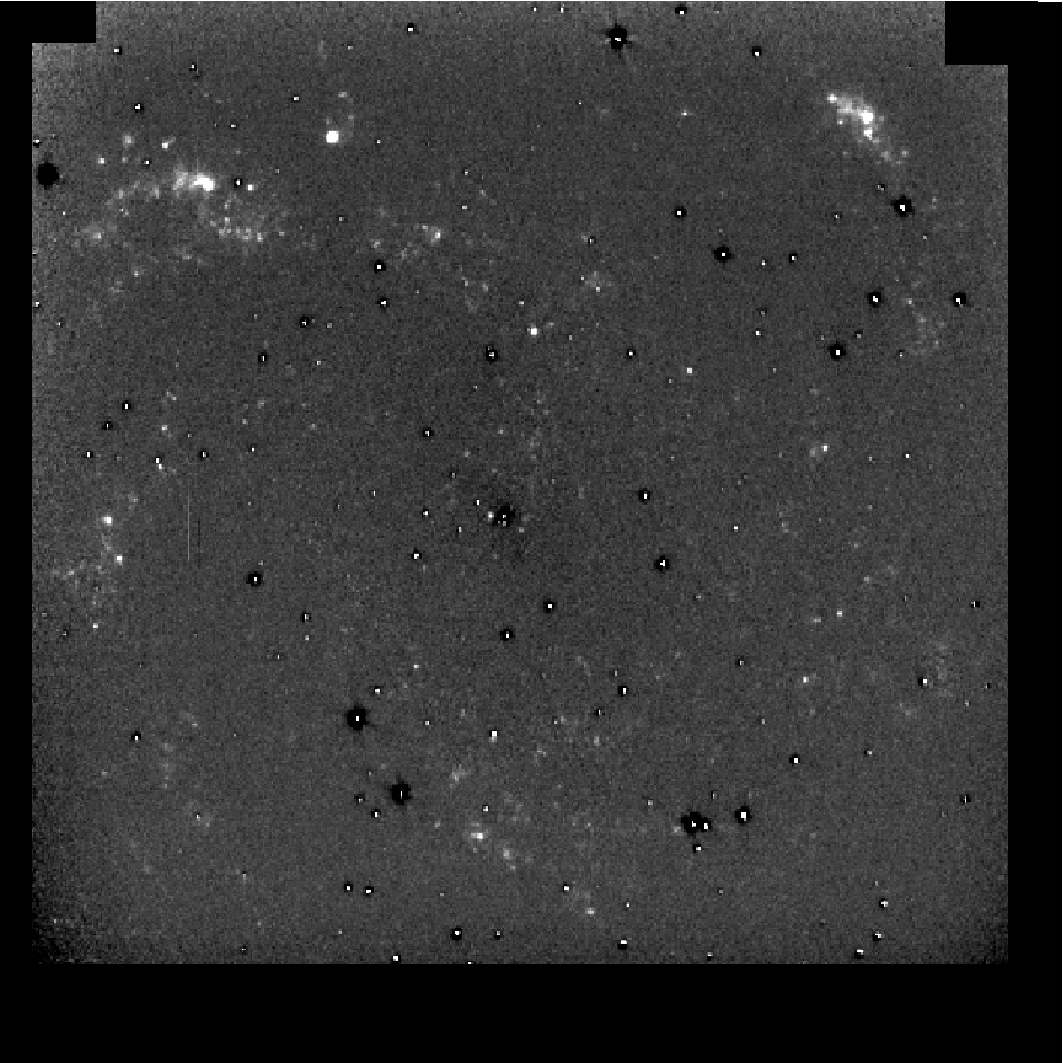} \\
 \end{tabular}
 \caption{Continuum substracted images for NGC~6946. Top left, H$\alpha$; top right, H$\beta$; botton left, [OII]; and bottom right [OIII]. North is at top and East at the left for all images. Scale as indicated.}
 \label{im69}
\end{figure*}

The continuum was substracted from each line image using the method described in Cedr\'es \& Cepa (2002). The calibrated and continuum substracted images from NGC~628 and NGC~6946 in all the lines are represented in Figs. \ref{im62} and \ref{im69} respectively.

\subsection{Corrections applied to the data}
To correct the emission lines from galactic and extragalactic extinction, we obtained first the $A_V$ value using the ratios of  the H$\alpha$ and H$\beta$ lines, assuming the case B for recombination, an electron density of 100\,cm$^{-3}$, a temperature of 10\,000\, K (James \& Puxley 1993), and the extinction law by Seaton (1979). Once we obtained $A_V$, employing the reddening law given by Shild (1977), we were able to determine the extinction for all the lines considered.

The underlying absorption due to massive stars in H~II regions and the emission from the underlying galactic disc in the H$\alpha$ and H$\beta$ continua were corrected employing the methods described in Cedr\'es \& Cepa (2002).

The H$\alpha$ fluxes were also corrected from the contribution of the [NII]$\lambda\lambda$6548,6584 lines employing the data from McCall et al. (1985).

\subsection{Flux extraction}
The H~II regions were detected and their fluxes were extracted employing the FOCAS task. To detect each region, we used the H$\alpha$ image (the one with the best signal to noise) and assumed that a region was detected when it had a flux per pixel 3$\sigma$ over the background. Then, following the method described in Cedr\'es \& Cepa (2002), a limiting isophote was created for each region with a flux 1$\sigma$ over the background. With those isophotes, we were able to obtain the fluxes in all the lines within the defined area in H$\alpha$, considering that a H~II region has emission in the selected line (different of H$\alpha$) when in has a flux 3$\sigma$ per pixel over the background. In total, we measured 209 H~II regions for NGC~628, and 226 H~II regions for NGC~6946.

\section{Results}
The studied physical parameters of the detected H~II regions for both galaxies are summarized in tables \ref{t1n62} to \ref{t2n69} for the first 10 regions of each galaxy. The complete tables are available electronically.

\begin{table*}
\caption{Positions and main properties of the first 10 H~II regions of NGC~628. The H$\alpha$ fluxes are extinction corrected. The complete table is available electronically.}
\begin{tabular}{c c c c c c c}
\hline
\hline
Number & X offset (arcsec) & Y offset (arcsec) & Deprojected R (arcmin) & F$_{\rm H\alpha}$ (10$^{14}$\,erg/cm$^{2}$/s) & A$_{V}$ (mag) & Area (arcsec$^{2}$)\\
\hline 
 1  &  87.49 &  183.08 &  3.390 & 14.784$\pm$1.445 & 0.00$\pm$0.06 & 36.21\\
 2  &   3.32 &  181.94 &  3.034 &  6.252$\pm$0.641 & 0.00$\pm$0.06 & 37.15\\
 3  &  14.15 &  176.81 &  2.958 &  1.555$\pm$0.184 & 0.00$\pm$0.07 &  9.93\\
 4  &  71.82 &  171.11 &  3.099 &  0.408$\pm$0.060 & 0.00$\pm$0.09 &  3.65\\
 5  & -43.51 &  153.92 &  2.666 &  7.855$\pm$0.791 & 0.00$\pm$0.06 & 18.59\\
 6  &  69.82 &  151.92 &  2.793 &  0.414$\pm$0.043 & 0.79$\pm$0.11 &  2.78\\
 7  & 167.96 &  145.75 &  3.725 &  1.613$\pm$0.184 & 0.07$\pm$0.07 &  8.30\\
 8  & 166.15 &  142.23 &  3.663 &  0.626$\pm$0.080 & 0.15$\pm$0.09 &  4.51\\
 9  & 145.25 &  138.15 &  3.357 &  1.158$\pm$0.142 & 0.00$\pm$0.08 &  8.38\\
10  & 165.77 &  136.34 &  3.596 &  0.866$\pm$0.095 & 0.32$\pm$0.08 &  6.46\\
\hline
\hline
\end{tabular}
\label{t1n62}
\end{table*}

\begin{table*}
\caption{Positions and main properties of the first 10 H~II regions of NGC~6946. The H$\alpha$ fluxes are extinction corrected. The complete table is available electronically.}
\begin{tabular}{c c c c c c c}
\hline
\hline
Number & X offset (arcsec) & Y offset (arcsec) & Deprojected R (arcmin) & F$_{\rm H\alpha}$ (10$^{14}$\,erg/cm$^{2}$/s) & A$_{V}$ (mag) & Area (arcsec$^{2}$)\\
\hline 
 1 & 67.36  & 187.27 & 3.824 & 0.570$\pm$0.079  &  0.09$\pm$0.09 &   4.95\\
 2 & 111.44 & 171.78 & 3.933 & 0.602$\pm$0.066  &  0.52$\pm$0.10 &   5.27\\
 3 & 127.02 & 145.18 & 3.680 & 69.807$\pm$4.167 &  0.66$\pm$0.06 & 196.64\\
 4 & -59.09 & 154.11 & 2.957 & 0.677$\pm$0.059  &  0.97$\pm$0.10 &   3.61\\
 5 & 114.86 & 151.55 & 3.642 & 0.725$\pm$0.058  &  1.14$\pm$0.11 &   3.21\\
 6 & 66.69  & 147.65 & 3.118 & 2.678$\pm$0.148  &  1.31$\pm$0.08 &   8.81\\
 7 & 36.96  & 146.89 & 2.900 & 3.431$\pm$0.212  &  1.01$\pm$0.08 &  13.10\\
 8 & -56.33 & 144.33 & 2.773 & 2.988$\pm$0.277  &  0.34$\pm$0.07 &  20.18\\
 9 & 116.47 & 143.76 & 3.538 & 1.305$\pm$0.089  &  1.16$\pm$0.09 &   6.53\\
10 & -62.79 & 138.72 & 2.705 & 6.209$\pm$0.332  &  1.12$\pm$0.07 &  10.22\\
\hline
\hline
\end{tabular}
\label{t1n69}
\end{table*}

\begin{table*}
\caption{Equivalent width, H$\beta$ flux and oxygen flux ratios, both extinction corrected, for the first 10 H~II regions of NGC~628. The complete table is available electronically.}
\begin{tabular}{c c c c c}
\hline
\hline
Number & log(EWH$\alpha$) (\AA) & F$_{\rm H\beta}$ (10$^{14}$\,erg/cm$^{2}$/s) & log ([OII]/H$\beta$) & log ([OIII]/H$\beta$)\\
\hline 
 1 &  2.496$\pm$0.071 & 6.044$\pm$1.351 & 0.501$\pm$0.099 & 0.259$\pm$0.098\\
 2 &  2.338$\pm$0.075 & 3.028$\pm$0.705 & 0.472$\pm$0.106 & 0.076$\pm$0.103\\
 3 &  2.521$\pm$0.088 & 0.641$\pm$0.171 & 0.479$\pm$0.139 & 0.118$\pm$0.123\\
 4 &  2.832$\pm$0.112 & 0.176$\pm$0.057 &     --          & 0.091$\pm$0.169\\
 5 &  0.531$\pm$0.073 & 4.801$\pm$0.993 &     --          &-1.203$\pm$0.110\\
 6 &  2.341$\pm$0.079 & 0.142$\pm$0.026 &     --          &       --       \\
 7 &  2.687$\pm$0.085 & 0.563$\pm$0.141 & 0.376$\pm$0.142 & 0.217$\pm$0.116\\
 8 &  2.528$\pm$0.095 & 0.218$\pm$0.059 &     --          & 0.132$\pm$0.138\\
 9 &  2.253$\pm$0.090 & 0.519$\pm$0.142 & 0.306$\pm$0.161 & 0.193$\pm$0.126\\
10 &  1.832$\pm$0.080 & 0.300$\pm$0.066 &    --           & 0.091$\pm$0.113\\
\hline
\hline
\end{tabular}
\label{t2n62}
\end{table*}

\begin{table*}
\caption{Equivalent width, H$\beta$ flux and oxygen flux ratios, both extinction corrected,  for the first 10 H~II regions of NGC~6946. The complete table is available electronically.}
\begin{tabular}{c c c c c}
\hline
\hline
Number & log(EWH$\alpha$) (\AA) & F$_{\rm H\beta}$ (10$^{14}$\,erg/cm$^{2}$/s) & log ([OII]/H$\beta$) & log ([OIII]/H$\beta$)\\
\hline 
 1 & 2.969$\pm$0.126 & 0.199$\pm$0.059 &        --        & -0.392$\pm$0.128\\
 2 & 2.860$\pm$0.101 & 0.208$\pm$0.043 & 0.535$\pm$ 0.090 & -0.430$\pm$0.092\\
 3 & 2.685$\pm$0.052 & 24.055$\pm$2.796& 0.446$\pm$ 0.051 & -0.092$\pm$0.050\\
 4 & 2.592$\pm$0.080 & 0.231$\pm$0.034 &        --        &        --       \\
 5 & 3.249$\pm$0.082 & 0.246$\pm$0.032 & 0.527$\pm$ 0.058 & -0.475$\pm$0.059\\
 6 & 3.113$\pm$0.054 & 0.904$\pm$0.084 & 0.494$\pm$ 0.040 & -0.012$\pm$0.040\\
 7 & 2.878$\pm$0.057 & 1.169$\pm$0.127 & 0.377$\pm$ 0.047 & -0.557$\pm$0.048\\
 8 & 2.544$\pm$0.081 & 1.041$\pm$0.196 & 0.337$\pm$ 0.082 & -0.524$\pm$0.082\\
 9 & 2.922$\pm$0.065 & 0.443$\pm$0.051 & 0.499$\pm$ 0.050 & -0.609$\pm$0.050\\
10 & 2.885$\pm$0.049 & 2.110$\pm$0.198 &        --        &        --       \\
\hline
\hline
\end{tabular}
\label{t2n69}
\end{table*}

\begin{figure*}
 \centering
 \begin{tabular}{c c}
 \resizebox{9cm}{!}{\includegraphics{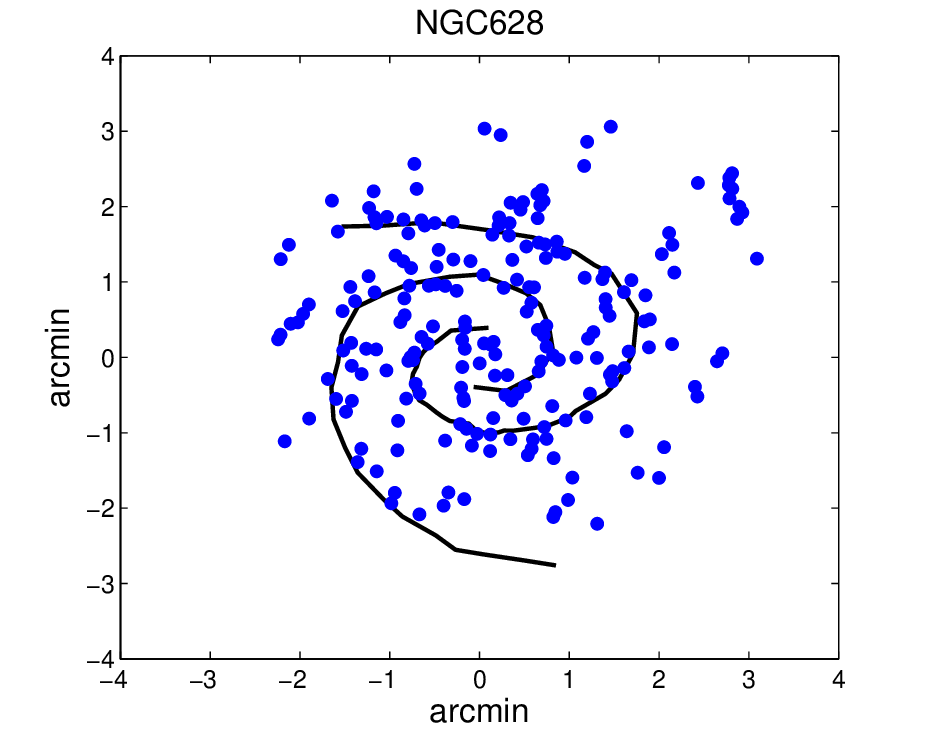}} & \resizebox{9cm}{!}{\includegraphics{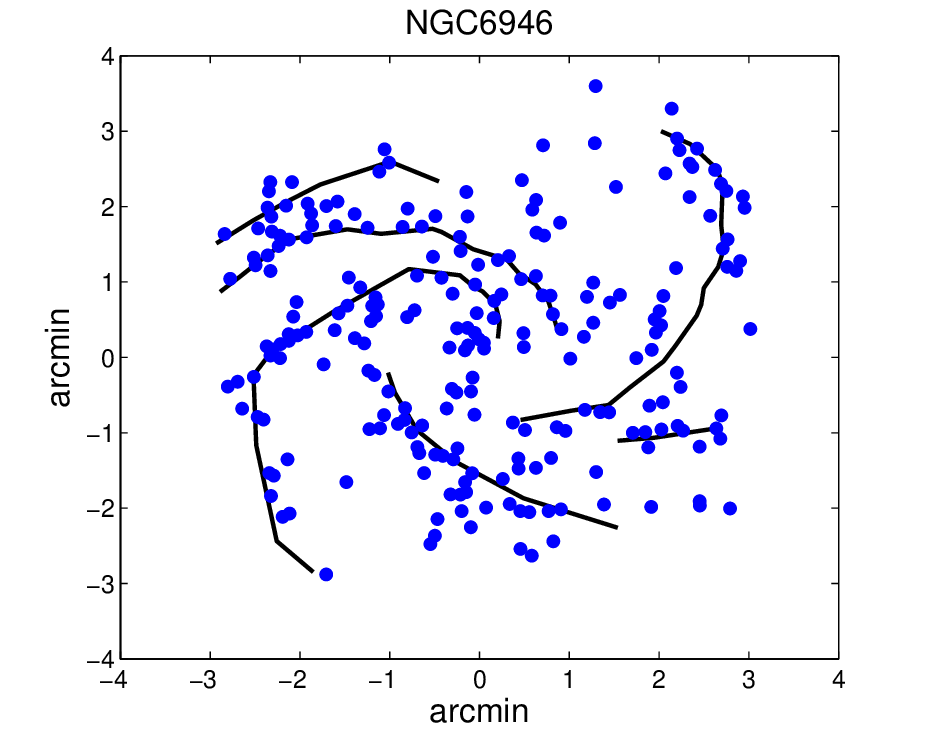}} \\
  \end{tabular}
 \caption{Positions of the H~II regions detected for NGC~628 (left panel) and NGC~6946 (right panel). The arms, as determined in Cedr\'es et al. (2012), are indicated by the continuous lines. North is at the top and East is at the left.}
 \label{tot}
\end{figure*}

In Fig. \ref{tot} the relative positions of the detected H~II regions for NGC~628 and NGC~6946, and also the positions of the arms, as determined in Cedr\'es et al. (2012), are shown.

\begin{figure*}
 \centering
 \begin{tabular}{c c}
 \resizebox{9cm}{!}{\includegraphics{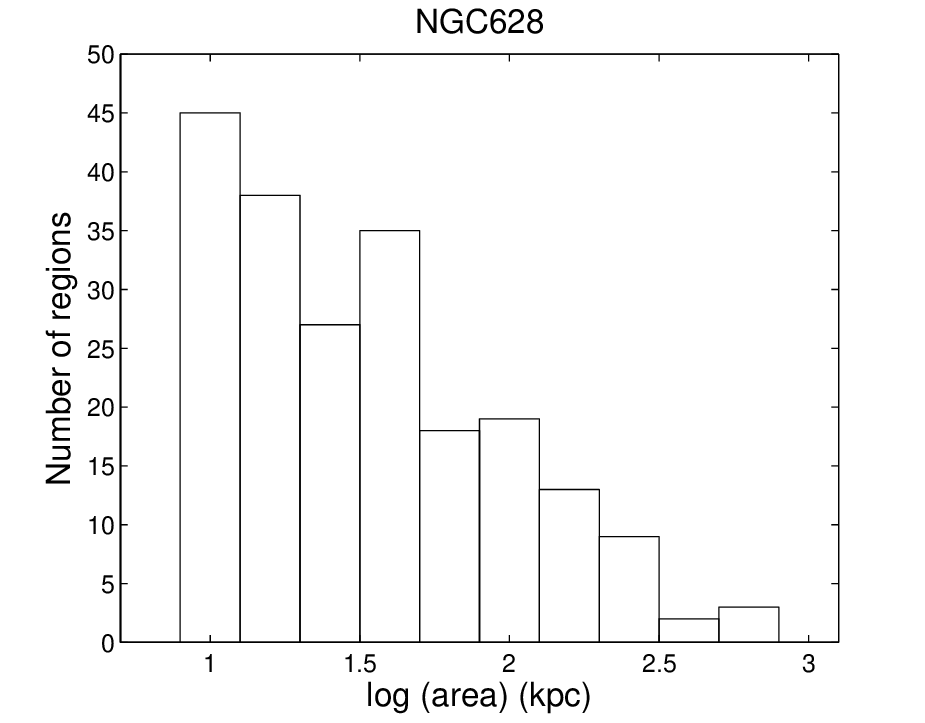}} & \resizebox{9cm}{!}{\includegraphics{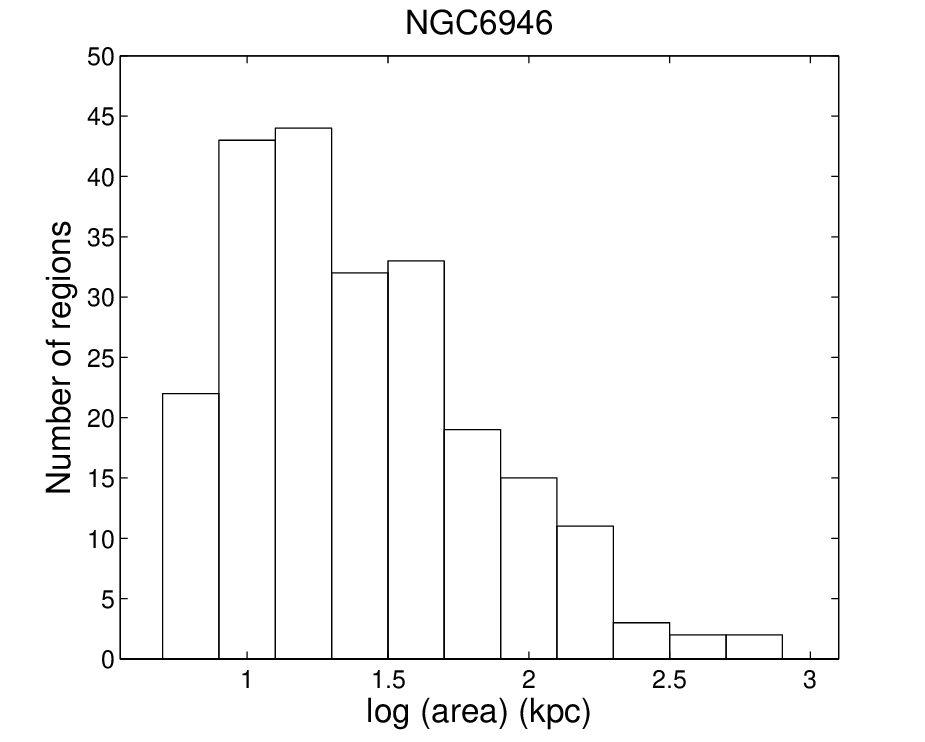}} \\
  \end{tabular}
 \caption{Histograms of the areas of the H~II regions in kpc, for NGC~628 (left panel) and NGC~6946 (right panel).}
 \label{area}
\end{figure*}

In Fig. \ref{area} we have represented the histograms of the area (in squared kpc) of the detected regions. For NGC~6946 there are more regions with an area less than 10\,kpc than for NGC~628. This is not surprising, because NGC~6946 is closer to us than NGC~628 and the seeing for both is similar, therefore, we are able to resolve smaller regions from images with the same depth for both galaxies.

\begin{figure*}
 \centering
 \begin{tabular}{c c}
 \resizebox{9cm}{!}{\includegraphics{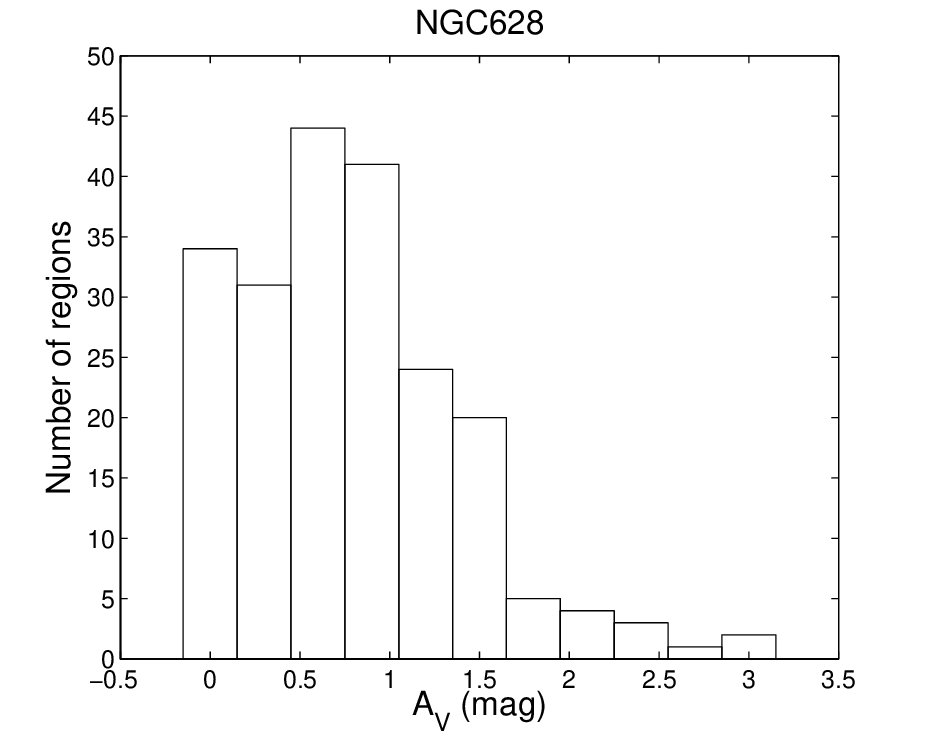}} & \resizebox{9cm}{!}{\includegraphics{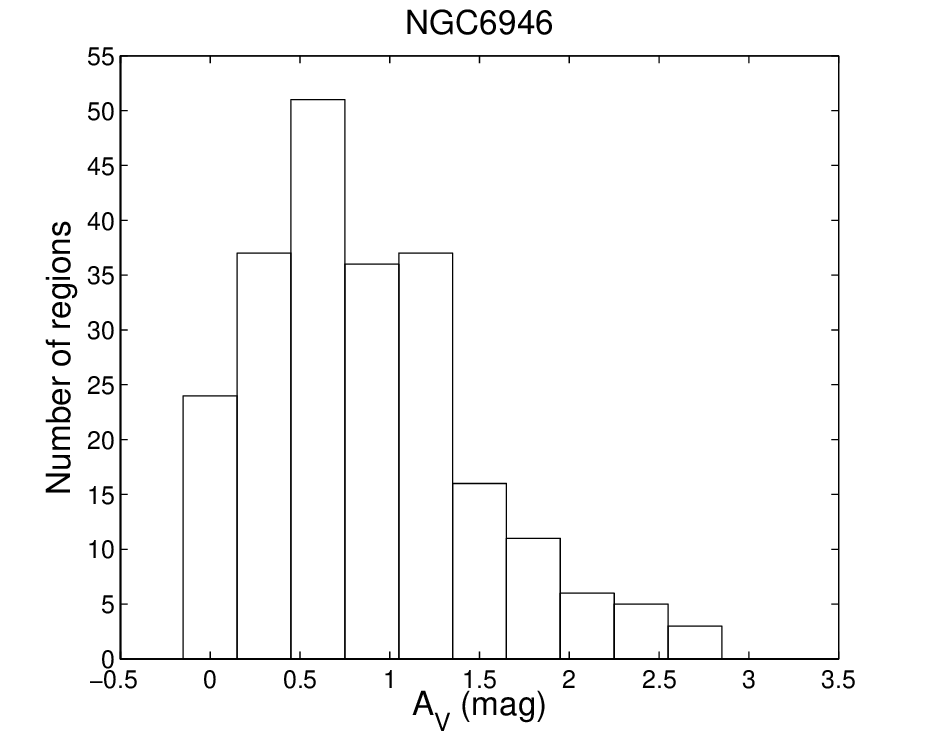}} \\
  \end{tabular}
 \caption{Histograms of the extinctions of the H~II regions in magnitudes, for NGC~628 (left panel) and NGC~6946 (right panel).}
 \label{hisext}
\end{figure*}

The extinction derived (represented as histograms in Fig. \ref{hisext}) for both galaxies is somewhat different from the values obtained by B\&R92. Our extinction is larger for NGC~628, with the maximum in the bin centered in A$_V$=0.9 (with a mean A$_V$=0.79). The maximum from B\&R92 is at the bin centered at A$_V$=0.125 (with a mean A$_V$=0.52). However, S\'anchez et al. (2011) give a mean value for A$_V$ between 1.24 and 1.04 (depending on the method employed: H~II regions or integrated spectrum, respectively), that is closer to our results. Moreover, S\'anchez et al. (2011) also find, like us, extinctions larger than A$_V$=2.5, while the larger extinction found in B\&R92 is about A$_V\simeq$1.6.
For NGC~6946, the derived value of the extinction for B\&R92 (A$_V$=1.01) is slighty larger than ours (A$_V$=0.9), however, the distribution and range for the values of the extinction are similar for both datasets.

\begin{figure*}
 \centering
 \begin{tabular}{c c}
  \resizebox{9cm}{!}{\includegraphics{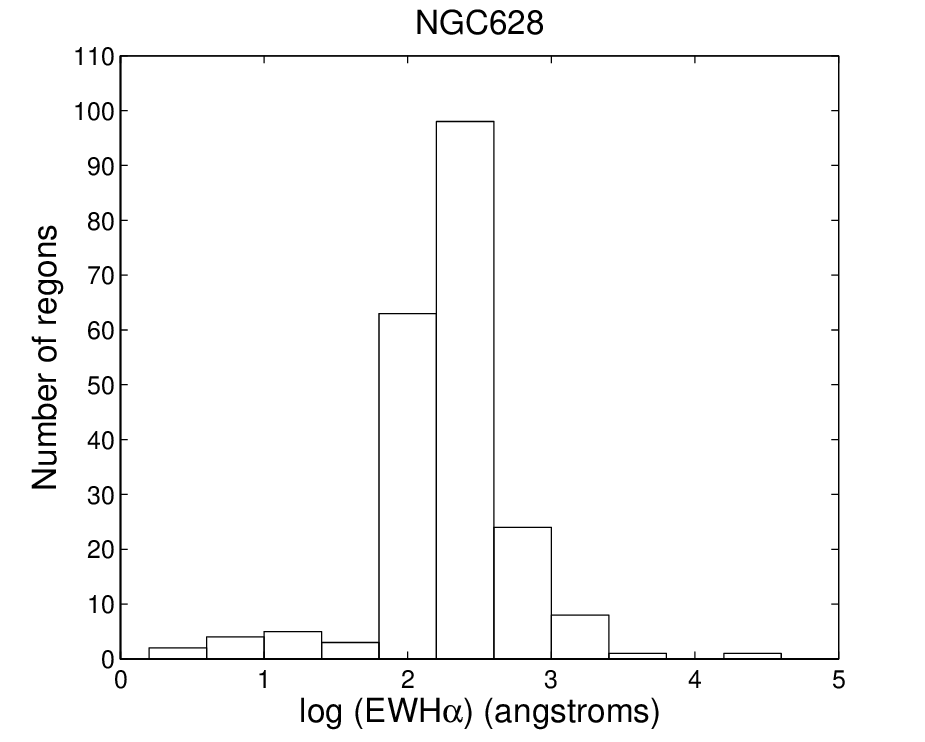}} & \resizebox{9cm}{!}{\includegraphics{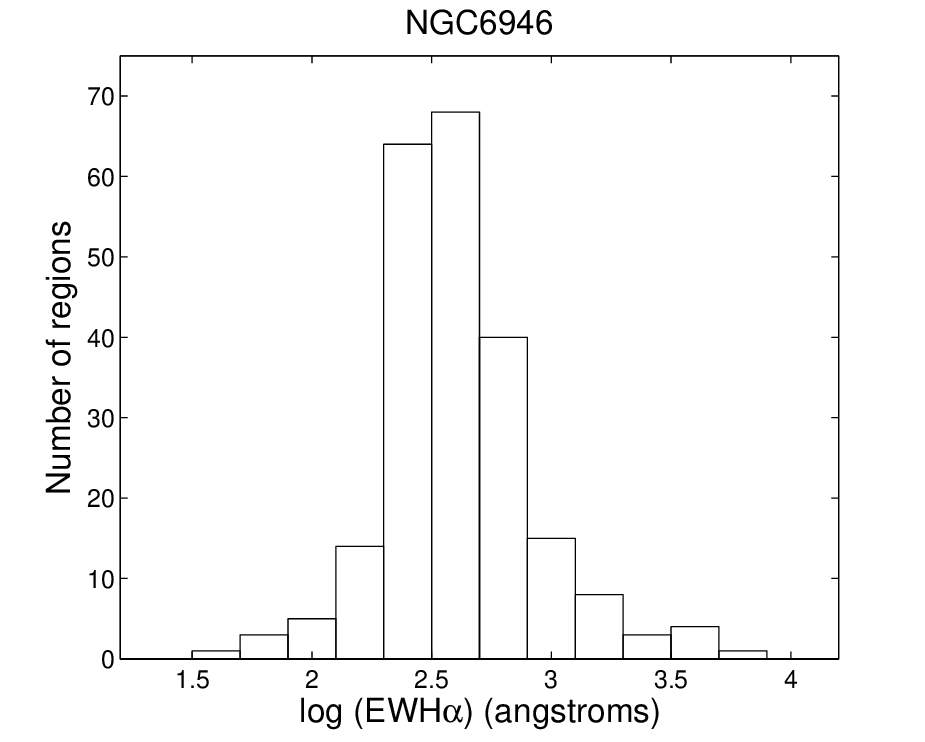}} \\
 \end{tabular}
 \caption{Histograms of the logarithm of the H$\alpha$ equivalent width for the H~II regions (in \"Angstroms), for NGC~628 (left panel) and NGC~6946 (right panel).}
 \label{histha}
\end{figure*}

\begin{figure*}
 \centering
 \begin{tabular}{c c}
  \resizebox{9cm}{!}{\includegraphics{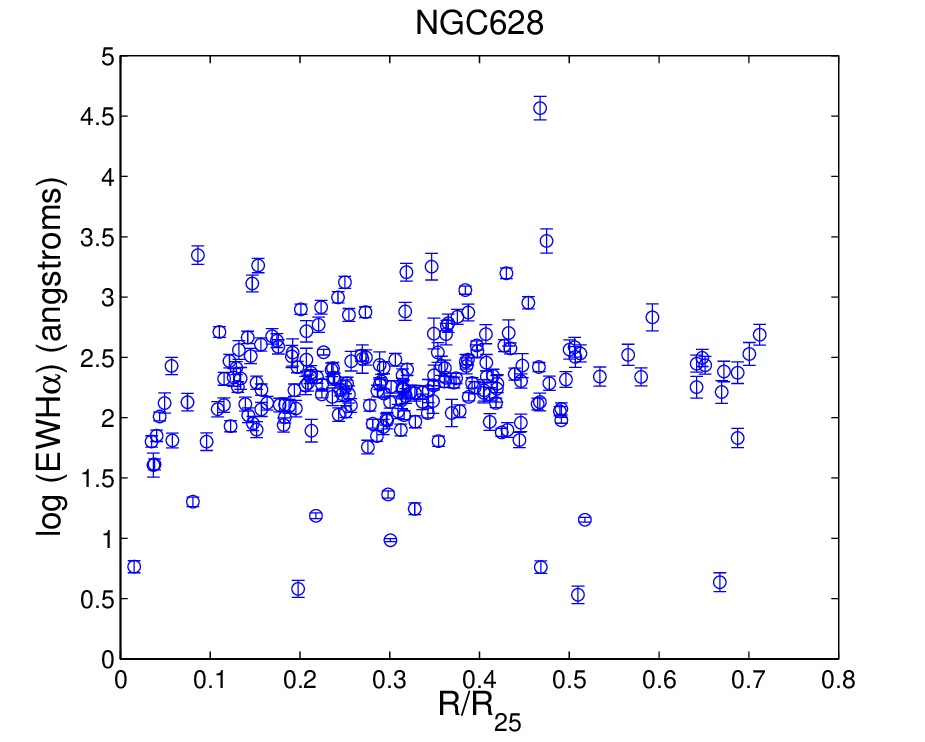}} & \resizebox{9cm}{!}{\includegraphics{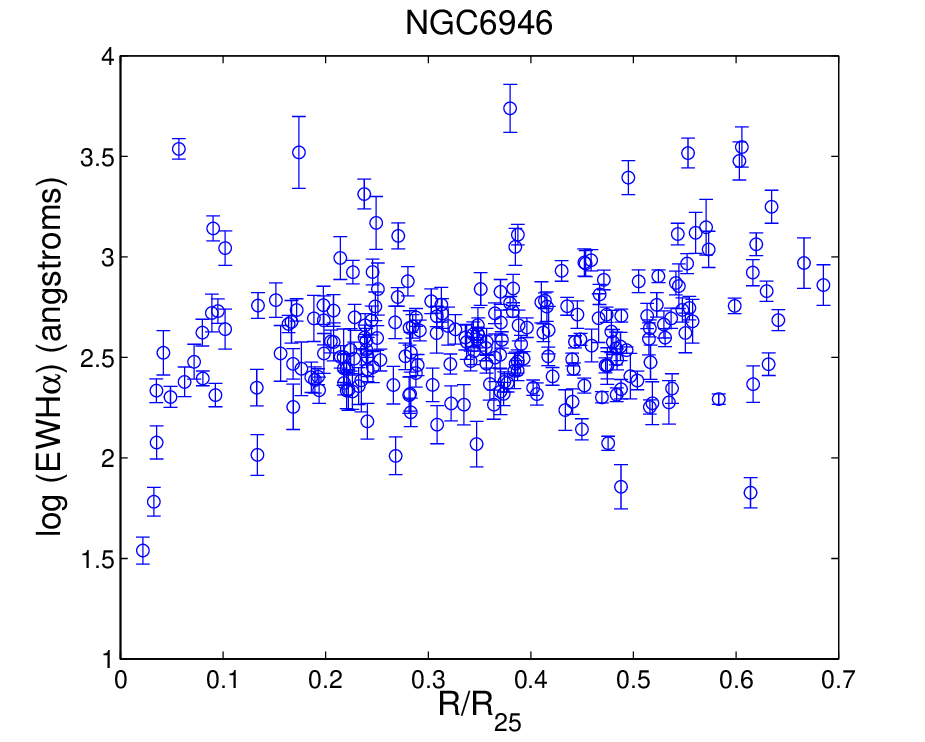}} \\
 \end{tabular}
 \caption{Logarithm of the H$\alpha$ equivalent width as a function of the R$_{25}$ radius for NGC~628 (left panel) and NGC~6946 (right panel).}
 \label{havsr}
\end{figure*}

In Fig. \ref{histha} the histograms for the logarithm of the H$\alpha$ equivalent width for NGC~628 (left) and NGC~6946 (right) are presented. The mean value of the H$\alpha$ equivalent width is slightly larger for NGC~6946, log(EWH$\alpha$)=2.6 than for NGC~628, log(EWH$\alpha$)=2.28. As discussed in Cedr\'es et al. (2011), this may be equivalent to mean older H~II regions for NGC~628, if we assume for both galaxies a Salpeter initial mass function.
From Fig. \ref{havsr}, where we represent the logarithm of the H$\alpha$ equivalent width as a function of the galactocentric radius for both galaxies, there is not a clear trend of the equivalent width with radius for the galaxies in the sample. A similar result was obtained in Cedr\'es \& Cepa (2002) for the galaxies NGC~5457 and NGC~4395.

\begin{figure*}
 \centering
 \begin{tabular}{c c}
  \resizebox{9cm}{!}{\includegraphics{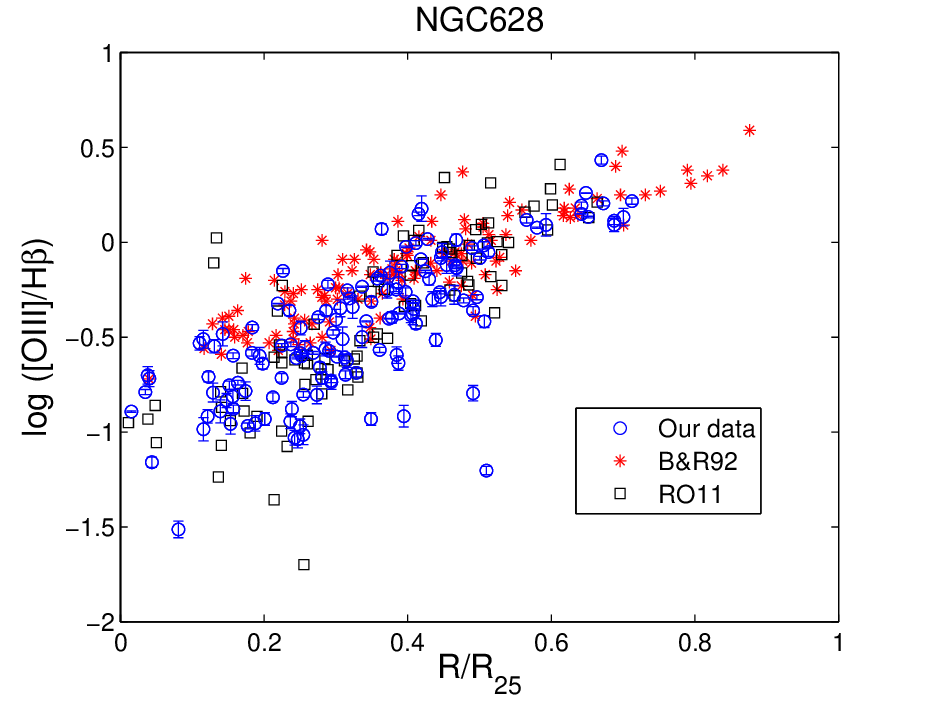}} & \resizebox{9cm}{!}{\includegraphics{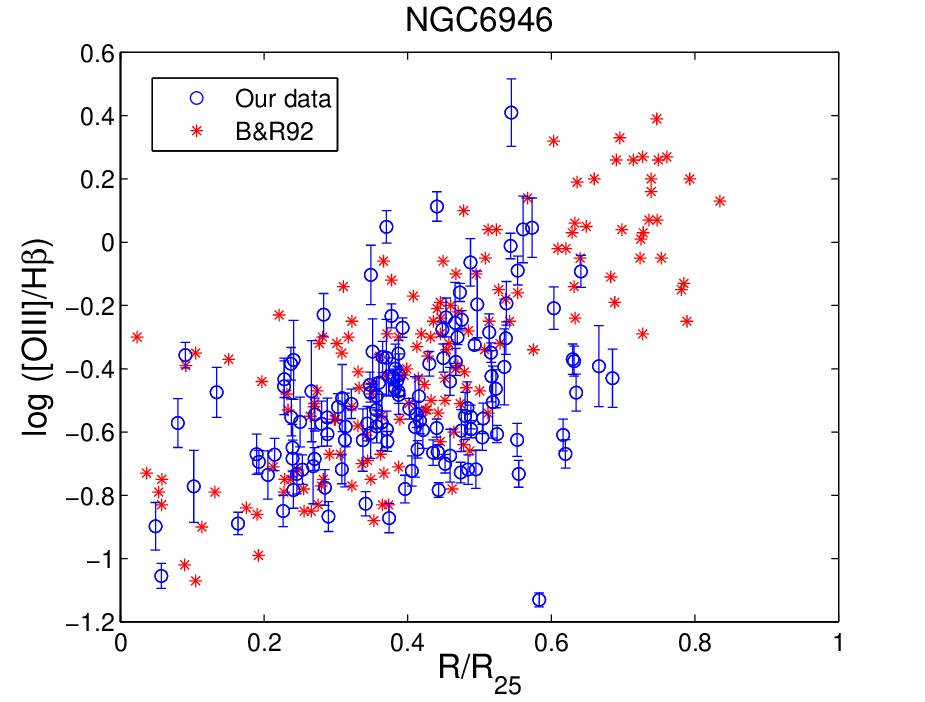}} \\
 \end{tabular}
 \caption{Logarithm of the [OIII]/H$\beta$ as a function of the R$_{25}$ radius for NGC~628 (left panel) and NGC~6946 (right panel). Our data are open circles, asterisks are the data from B\&R92 and open squares are the data from RO11.}
 \label{o3}
\end{figure*}

In Fig. \ref{o3} we have represented the logarithm of the ratio [OIII]$\lambda\lambda$4959,5007/H$\beta$ as a function of the galactocentric radius. When comparing both galaxies with data from the literature, there is a clear agreement with our results, even for data obtained using different techniques, like the RO11 data for NGC~628. The [OIII]/H$\beta$ ratio increases with galactocentric distances for both galaxies, which indicates the presence of an oxygen abundance gradient for both galaxies. The data from B\&R92 covers a slighty larger field for both galaxies than our data, so more regions are presented over R/R$_{25}$$>$0.75, and those are the ones with mean larger values for [OIII]$\lambda\lambda$4959,5007/H$\beta$.

\begin{figure*}
 \centering
 \begin{tabular}{c c}
  \resizebox{9cm}{!}{\includegraphics{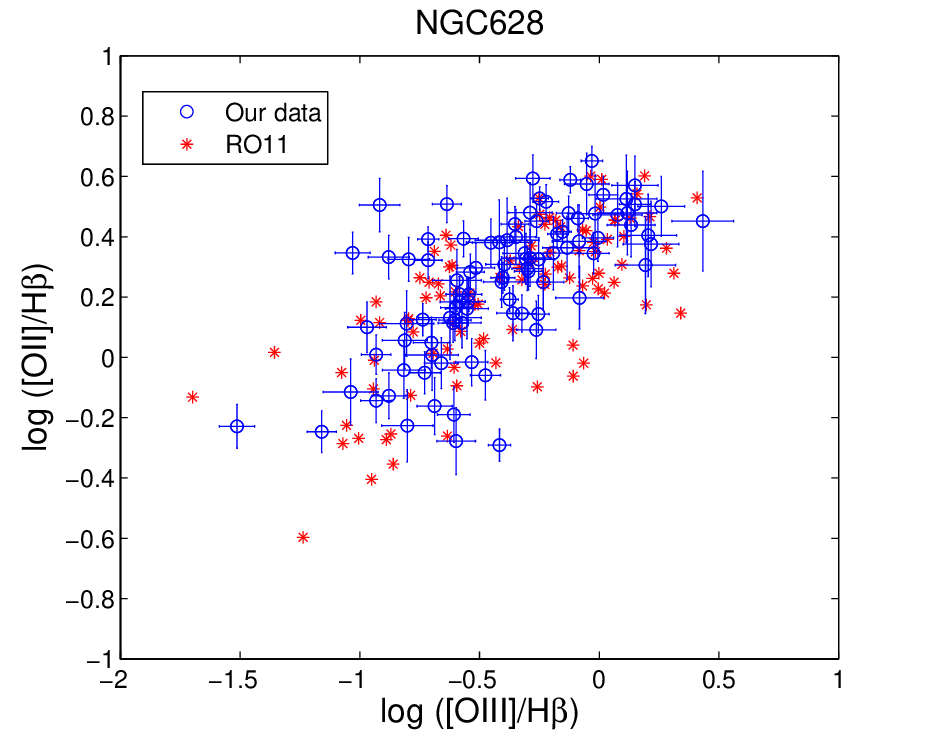}} & \resizebox{9cm}{!}{\includegraphics{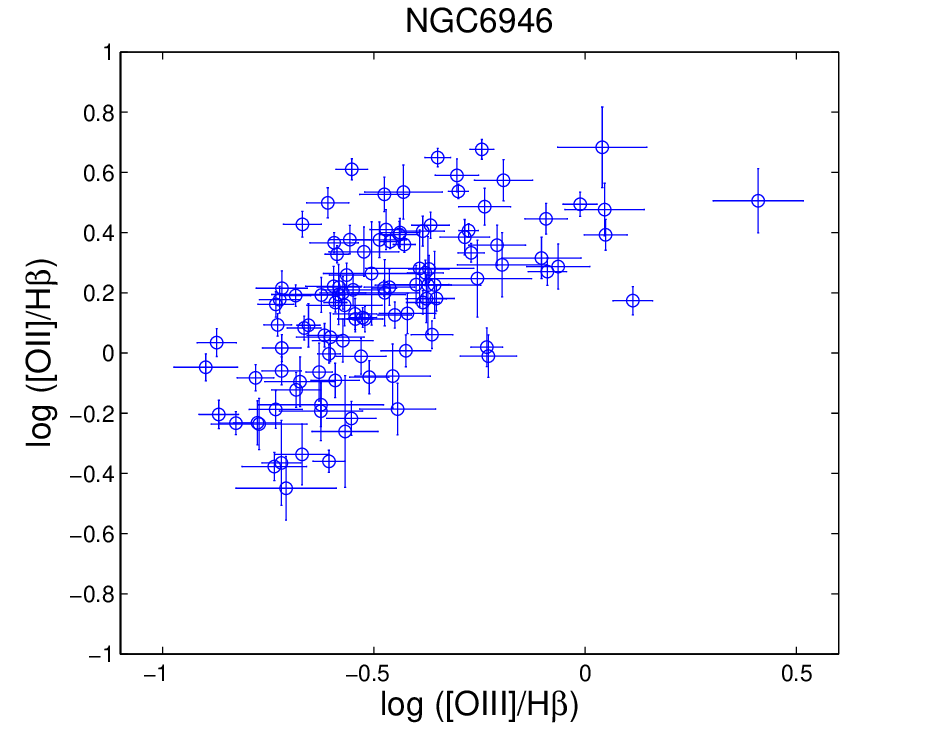}} \\
 \end{tabular}
 \caption{Logarithm of the [OII]/H$\beta$ versus of the logarithm of [OIII]/H$\beta$ for NGC~628 (left panel) and NGC~6946 (right panel). Our data are open circles, asterisks are the data from RO11.}
 \label{o2o3}
\end{figure*}

The relationship between the logarithm of the [OII]/H$\beta$ and [OIII]/H$\beta$ ratios is represented in Fig. \ref{o2o3} for NGC~628 and NGC~6946. For both galaxies the data is distributed in the expected locus, and for NGC~628 there is a very good agreement with RO11 data. However, for NGC~6946, the excitation seeems to be somewhat lower (at least in the parts of the galaxy where we have data) when compared with NGC~628.

\section{Determination of the oxygen abundance}
\subsection{The problem with empirical calibrations}
The main method employed in order to derive the metallicity from galactic H~II regions is through electron temperature sensitive lines, the so-called direct method (Searle 1971, Garnett \& Kennicutt 1994, Kennicutt et al. 2003, Izotov et al. 2006, among others). However, the oxygen line [OIII]$\lambda$4363, critical to determine the electron temperature (Osterbrock \& Ferland, 2006 or Izotov et al. 2006), is weak and difficult to detect in the inner parts of spiral galaxies, where the oxygen abundance is high.
So, in order to determine the metallicity with nebular lines, we need alternative methods based on empirical calibrations. From the literature, several calibrations have been proposed: Edmunds \& Pagel (1984), McCall et al. (1985), Zaritsky et al. (1994), Kewley \& Dopita (2002) and Kobulnicky \& Kewley (2004), based on the $R_{23}$ parameter, defined as
\begin{equation}
R_{23}=\frac{[{\rm OII}]\lambda\lambda3727,29+[{\rm OIII}]\lambda\lambda4959,5007}{{\rm H}\beta}
\end{equation}
Or Pilyugin (2001a, b) and Pilyugin \& Thuan (2005), that involves the $R_{23}$ parameter and the excitation $P$, defined as
\begin{equation}
P=\frac{[{\rm OIII}]\lambda\lambda4959,5007}{[{\rm OII}]\lambda\lambda3727,29+[{\rm OIII}]\lambda\lambda4959,5007}
\label{pp}
\end{equation}
or even van Zee et al. (1998), Denicol\'o et al. (2002) and Pettini \& Pagel, with the parameter $N_2$.
The data set presented here allows us to obtain the $R_{23}$ and $P$ parameters only. However, it has to be pointed out that the $R_{23}$ calibrations provide oxygen abundance determinations about 0.5\,dex higher than the ones obtained by the direct method (L\'opez-S\'anchez \& Esteban, 2010, Cedr\'es et al. 2004). According to L\'opez-S\'anchez \& Esteban (2010), the Pilyugin \& Thuan (2005) (thereafter PT05) calibration is the most suitable to derive oxygen abundances, because is the one in better agreement with the metallicity obtained by the direct method. On the other hand, Moustakas et al. (2010) argue that the scarcity of calibration points with high oxygen abundance and low excitation used in the development of the PT05 method may cause some problems in the metallicity obtained, an effect that is recognized also in PT05. Moreover, Moustakas et al. (2010), also recommend employing two calibrations for SINGS galaxies, Kobulnicky \& Kewley (2004) (thereafter KK04) and PT05, or an average of them, in order to diminish the possible systematic errors introduced by each one.

\begin{figure}
 \centering
   \resizebox{\hsize}{!}{\includegraphics{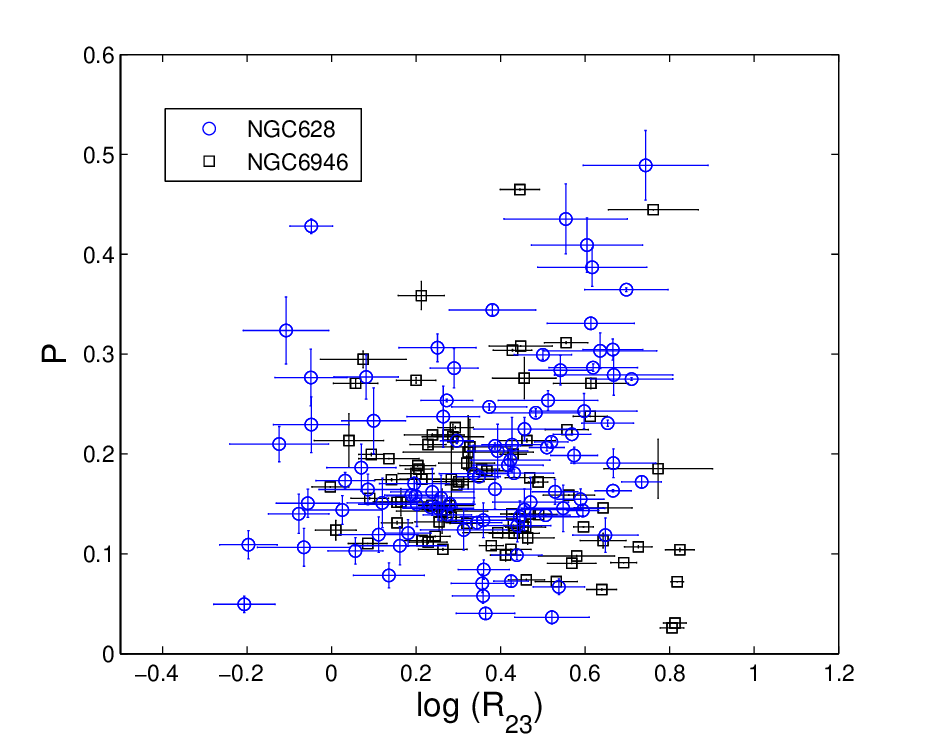}}
 \caption{Excitation $P$ (Eq. \ref{pp}) as a function of the logarithm of $R_{23}$ for NGC~628 (open circles) and NGC~6946 (open squares).}
 \label{pvsr23}
\end{figure}

In Fig. \ref{pvsr23} we have represented for NGC~628 and NGC~6946, the $P$ parameter as a function of the logarithm of $R_{23}$. Almost all our regions have $P<0.4$, so that they are located within a range of $P$ that it is not well sampled, because for PT05 the larger number of calibration points are in $P>0.6$ for the upper branch of the metallicity. 
However, in Cedr\'es et al. (2004), it is shown that even for M~101 regions with low value of $P$, the calibration from Pilyugin (2001a,b) (a previous version of the present PT05) shows a more than reasonable agreement with the metallicity obtained by the direct method. This seems to indicate that for low values of $P$, PT05 may present more scatter, but that the mean oxygen abundance value determined will behave as good as that for larger values of $P$, where more calibration points are present.

The metallicity is derived, following PT05 method, for the upper branch using:
\begin{equation}
12+\log({\rm O/H})_{up}=\frac{R_{23}+726.1+842.2P+337.5P^2}{85.96+82.76P+43.98P^2+1.793R_{23}}
\label{uppil}
\end{equation}
And for the lower branch using:
\begin{equation}
12+\log({\rm O/H})_{low}=\frac{R_{23}+106.4+106.8P-3.40P^2}{17.72+6.60P+6.95P^2-0.302R_{23}}
\label{lowpil}
\end{equation}
To obtain the metallicity using the KK04 calibration, we apply the equations
\begin{eqnarray}
12+\log({\rm O/H})_{up}=9.72-0.777x-0.951x^2-0.072x^3-0.811x^4- \nonumber \\
\log(q)(0.0737-0.0713x-0.141x^2+0.0373x^3-0.058x^4)
\end{eqnarray}
\begin{eqnarray}
12+\log({\rm O/H})_{low}=9.40+4.65x-3.17x^2- \nonumber \\
\log(q)(0.272+0.547x-0.513x^2)
\end{eqnarray}
where $x=\log(R_{23})$ and $q$ is given by
\begin{eqnarray}
\log(q)=32.81-1.153y^2+z(-3.396-0.025y+0.1444y^2)\times \nonumber \\
\left[4.603-0.3119y-0.163y^2+z(-0.48+0.0271y+0.02037y^2)\right]^{-1}
\end{eqnarray}
where $z=12+\log({\rm O/H})$ and 
\begin{equation}
y=\log\left(\frac{{\rm [OIII]}\lambda\lambda4959,5007}{{\rm [OII]}\lambda3727}\right)
\end{equation}

In this study, we will obtain the metallicity employing PT05 method and we will compare it with the results from KK04 and, when avaliable, metallicity values obtained through the direct method.
In tables \ref{meta628} and \ref{meta6946} the values for the parameters $P$ and $R_{23}$, and the oxygen abundance derived using the PT05 calibration for the first 10 H~II regions of NGC~628 and NGC~6946, respectively, are presented.

\begin{table*}
\caption{Values for the parameters $P$ and $R_{23}$, and the metallicity derived through PT05 method for the first 10 H~II regions of NGC~628. The complete table is available electronically.}
\begin{tabular}{c c c c}
\hline\hline
Number & $P$ & $\log R_{23}$ & 12+log(O/H)\\
\hline
 1 & 0.365 & 0.697 & 8.273$\pm$ 0.024\\
 2 & 0.287 & 0.619 & 8.276$\pm$ 0.041\\
 3 & 0.303 & 0.636 & 8.277$\pm$ 0.163\\
 4 &   --  &   --  &   -- \\
 5 &   --  &   --  &   -- \\
 6 &   --  &   --  &   -- \\
 7 & 0.409 & 0.605 & 7.978$\pm$ 0.222\\
 8 &   --  &   --  &   -- \\
 9 & 0.435 & 0.554 & 7.983$\pm$ 0.271\\
10 &   --  &   --  &   -- \\
\hline\hline
\end{tabular}
\label{meta628}
\end{table*}

\begin{table*}
\caption{Values for the parameters $P$ and $R_{23}$, and the metallicity derived through PT05 method for the first 10 H~II regions of NGC~6946. The complete table is available electronically.}
\begin{tabular}{c c c c}
\hline\hline
Number & $P$ & $\log R_{23}$ & 12+log(O/H)\\
\hline
 1  &	 --  &   --  &   --\\
 2  &  0.098 & 0.580 &   8.052 $\pm$0.128\\
 3  &  0.224 & 0.557 &   8.266 $\pm$0.063\\
 4  &	 --  &   --  &   --\\
 5  &  0.090 & 0.568 &   8.052 $\pm$0.081\\
 6  &  0.238 & 0.612 &   8.223 $\pm$0.055\\
 7  &  0.104 & 0.424 &   8.216 $\pm$0.049\\
 8  &  0.121 & 0.393 &   8.268 $\pm$0.077\\
 9  &  0.072 & 0.532 &   8.061 $\pm$0.063\\
10  &	 --  &   --  &   --\\
\hline\hline
\end{tabular}
\label{meta6946}
\end{table*}

\subsection{The metallicity of NGC~628}

\begin{figure}
 \centering
   \resizebox{\hsize}{!}{\includegraphics{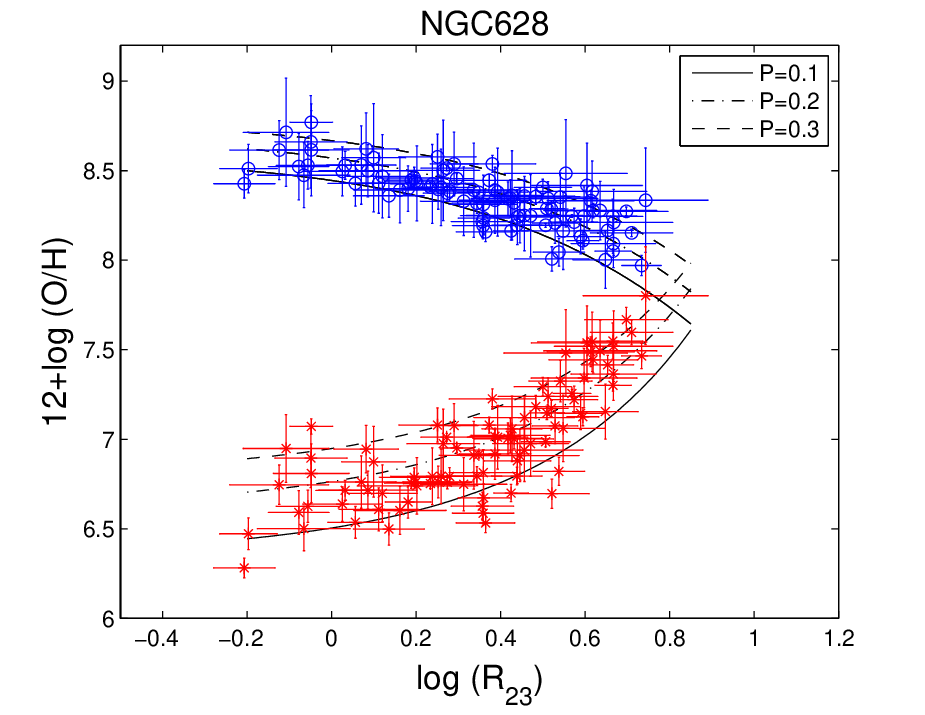}}
 \caption{Oxygen abundance for NGC~628 as a function of the logarithm of $R_{23}$ for different values of $P$. Open circles are the regions with the metallicity derived using the upper branch calibration. Asterisk are the regions with metallicity derived using the lower branch calibration.}
 \label{metvsr23628}
\end{figure}

From Fig. \ref{pvsr23}, using PT05 calibration, we can observe that most regions detected for NGC~628 are in the $0.1<P<0.3$ regime. In PT05, the authors situated the transition zone between the upper and the lower branch of the calibration aproximately between $8.0<12+\log(\rm{O/H})<8.3$. However, this is generally a zone populated by regions with a large range of values of $P$, and with most of the regions with $P<0.5$. Taking into account that our values of $P$ are more restricted, that the transition zone for low values of $P$ is below $12+\log(\rm{O/H})=8.0$, and that we are only sampling the central parts of the galaxy, where the oxygen abundance is moderately high, we can assume that all the regions with $8.0<12+\log(\rm{O/H})<8.3$ belong to the upper branch of the metallicity calibration (Fig. \ref{metvsr23628}). 

\begin{figure}
 \centering
   \resizebox{\hsize}{!}{\includegraphics{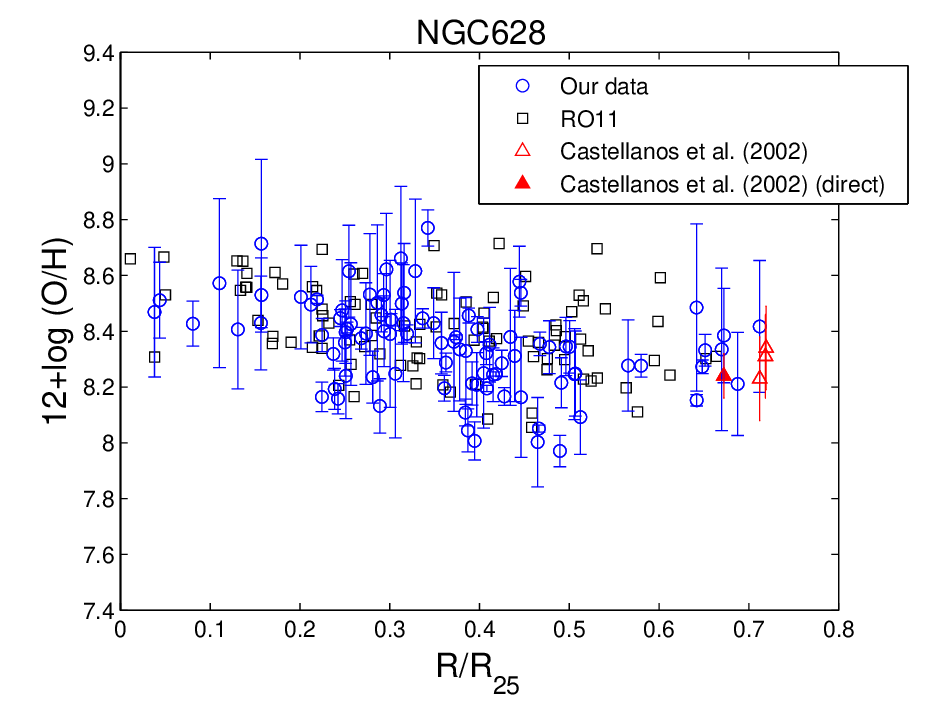}}
 \caption{Oxygen abundance for NGC~628 as a function of the galactocentric radius, employing the calibration from PT05. Our data are the open circles. The open squares are the data from RO11 and the open and filled triangles are the data from Castellanos et al. (2002).}
 \label{metvsr628}
\end{figure}

In Fig. \ref{metvsr628} we show the oxygen abundance obtained through PT05 calibration for the H~II regions of NGC~628 versus galactocentric radius. Again, there is a reasonable agreement between our data and RO11 data. We have also represented data from Castellanos et al. (2002) (filled triangle, region H13), where the oxygen abundance was obtained employing the direct method, or the regions where an empirical $S_{23}$ calibration was used (open triangles). If we fit our data, employing the least squares method, we obtain a gradient with the galactocentric radius of -0.36\,dex/R$_{25}$, very similar to the one obtained with the RO11 data, of -0.38\,dex/R$_{25}$.

\begin{figure}
 \centering
   \resizebox{\hsize}{!}{\includegraphics{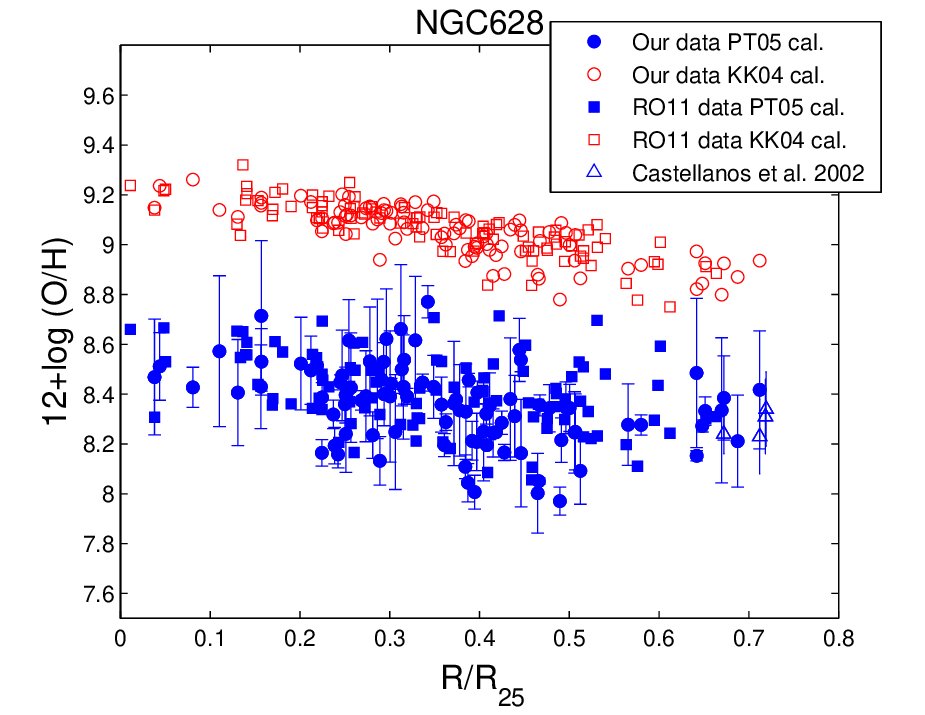}}
 \caption{Oxygen abundance for NGC~628 as a function of the galactocentric radius, employing the calibration from PT07 (filled circles and squares) and KK04 calibration (open circles and squares). Data from Castellanos et al.(2002) are the open triangles.}
 \label{metco628}
\end{figure}

In Fig. \ref{metco628} we have the metallicity for NGC~628 as in Fig. \ref{metvsr628} but, to serve as comparison, we have included the metallicity obtained through KK04 calibration for our data (open circles) and those of RO11 (open squares).

It is clear that our data follows closely the same behaviour as RO11 data. There is also a noticeable difference between the oxygen abundances yielded for both calibrations, as large as 0.6\,dex in the inner parts of the galaxy.
As expected, the KK04 calibration presents a tighter correlation with galactocentric radius while PT05 shows a larger dispersion. This is because most of the H~II regions in the inner parts of NGC~628 present $P<0.4$ and, as mentioned previously, this range is not well sampled by the PT05 calibration.
However, if we take into account the position of the regions from Castellanos et al. (2002) in the figure, it is clear that they lay closer to the gradient drawn by the PT05 calibration than the one from KK04. This is in agreement with the results from Cedr\'es et al. (2004) and L\'opez-S\'anchez \& Esteban (2010). 

\begin{table*}
\caption{Radial oxygen abundance gradients and central oxygen abundances for NGC~628}
\begin{tabular}{c |c c c c}
\hline\hline
 & \multicolumn{2}{c}{Gradient (dex/R$_{25}$)} & \multicolumn{2}{c}{Central metallicity}\\
 & PT05 & KK04 & PT05 & KK04\\
This work & $-$0.36$\pm0.1$ & $-$0.57$\pm0.1$ & 8.51$\pm0.04$ & 9.25$\pm0.04$\\
Moustakas et al. (2010) & $-$0.27$\pm0.05$ & $-$0.57$\pm0.04$ & 8.43$\pm0.02$ & 9.19$\pm0.02$\\
RO11 & $-$0.38$\pm0.02$ & $-$0.68$\pm0.03$ & 8.54$\pm0.01$ & 9.31$\pm0.02$\\
\hline\hline
\end{tabular}
\label{grad628}
\end{table*}

In table \ref{grad628} it is shown a summary of the gradients (expressed as dex/R$_{25}$) and the central oxygen abundances, obtained for NGC~628 using both calibrations and three different datasets: Moustakas et al. (2010), RO11, and our data.
There is an overall agreement between the three datasets within the uncertainties, for both calibrations. We must stress that Moustakas et al. (2010) data was obtained through long slit spectroscopy, and that RO11 data was obtained using integral field spectroscopy methods, while our data was obtained employing narrow-band imaging. This implies that narrow-band imaging techniques are, at least, as accurate and less time consuming as any spectroscopic method (long slit or integral field).

\begin{figure*}[h!]
 \centering
 \begin{tabular}{c c} 
   \resizebox{9cm}{!}{\includegraphics{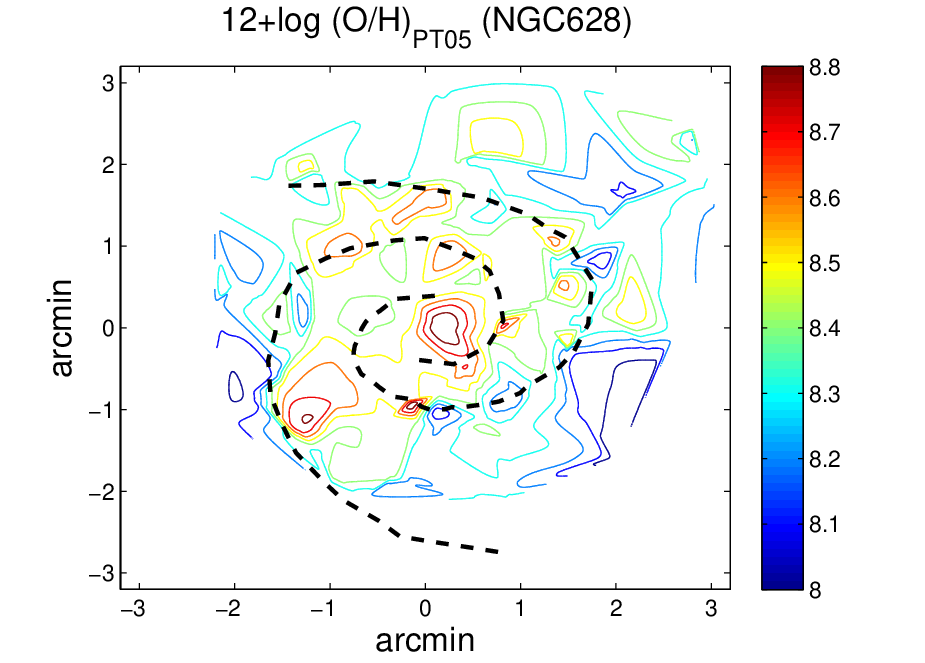}} & \resizebox{9cm}{!}{\includegraphics{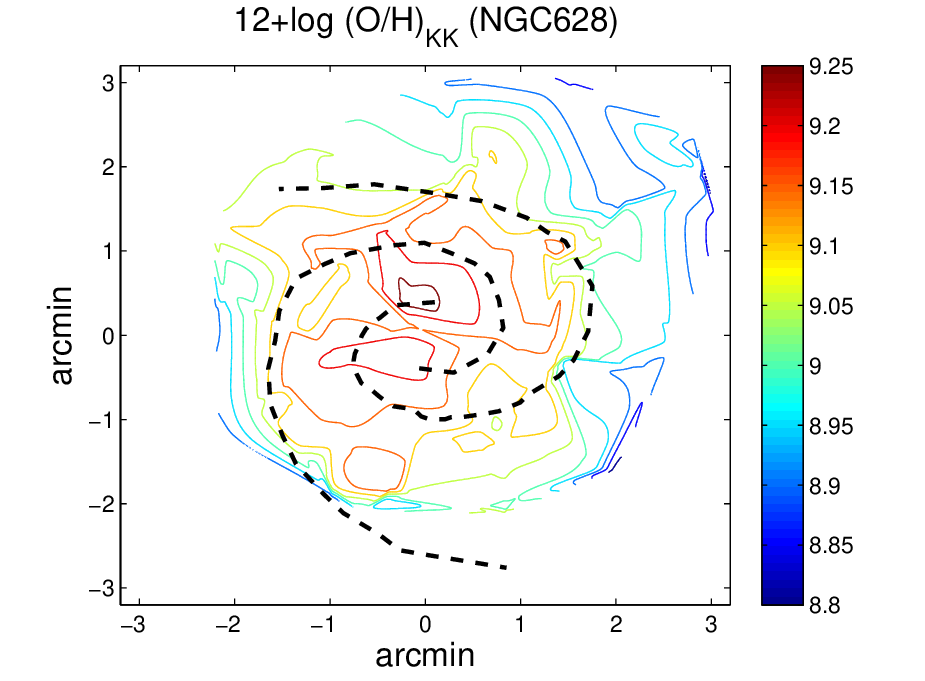}}\\
 \end{tabular}
 \caption{Contour plot of the oxygen abundance, determined using the PT05 calibration (left panel) and the KK04 calibration (right panel), for NGC~628. North is at the top and East is at the left. Spiral arms are indicated by the dashed lines (Cedr\'es et al. 2012).}
 \label{contmp628}
\end{figure*}

To explore the azimuthal behaviour of the oxygen abundance, we created a two dimensional continuous model of the metallicity from our discrete H~II regions, using a Delaunay triangulation (Barber et al. 1996). In order to have enough datapoints to make the required interpolations, we have included in the grid all the regions detected, even those with fluxes below the 3$\sigma$ criterium for [OII] and [OIII]. In Fig. \ref{contmp628} the metallicity contours of NGC~628, using the PT05 calibration (left panel) and KK04 calibration (right panel), are shown.
The results from the PT05 calibration are similar to the ones obtained by S\'anchez et al. (2011). There are several knots with larger metallicity values that seem to be associated to the arms. However, those knots dissapear for the KK04 calibration, where a more smoother and clearer gradient is shown instead.

\begin{figure*}[h!]
 \centering
 \includegraphics{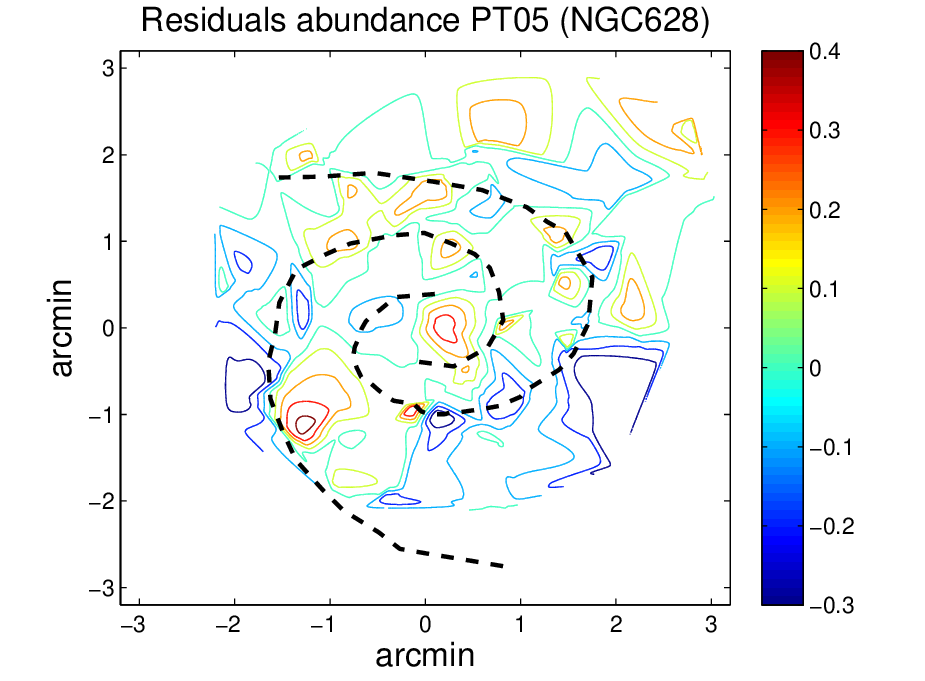}
 \caption{Contour plot of the residuals of the oxygen abundance, determined using the PT05 calibration, after the subtraction of the mean radial metallicity gradient, for NGC~628. North is a the top and East is at the left. Spiral arms are indicated by the dashed lines (Cedr\'es et al. 2012).}
 \label{res628}
\end{figure*}

\begin{figure*}[h!]
 \centering
 \includegraphics{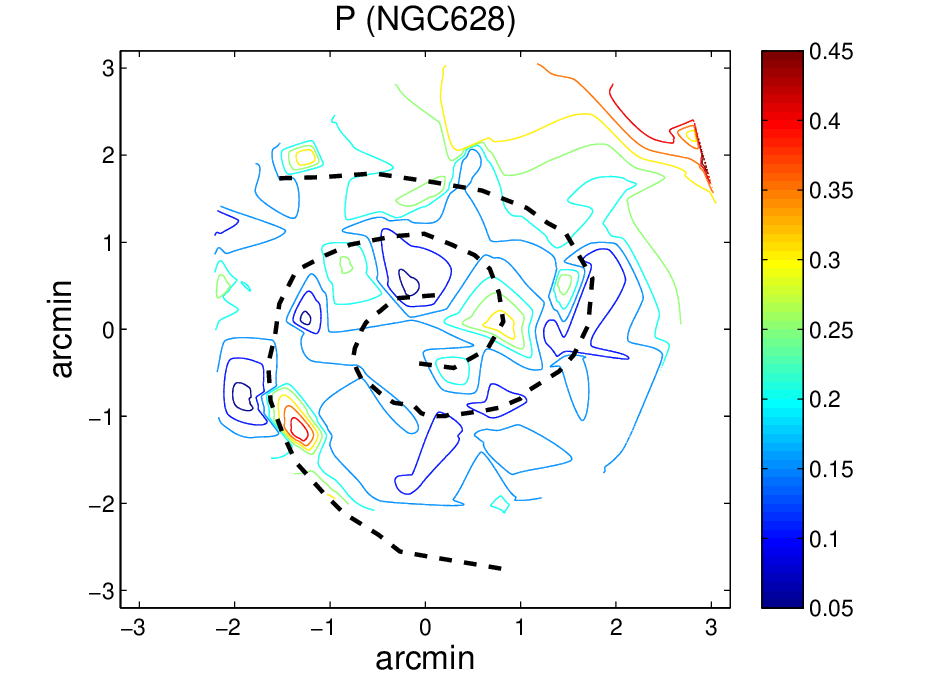}
 \caption{Controur plot of the parameter $P$ for NGC~628. North is at the top and East is at the left. Spiral arms are indicated by the dashed lines (Cedr\'es et al. 2012).}
 \label{contp628}
\end{figure*}

To highlight the high metallicity knots and the possible azimuthal variations, we have subtracted the radial metallicity gradient from the disc. The residuals are shown in Fig. \ref{res628}. 

Taking into account that the main difference between both calibrations (PT05 and KK04) is the inclusion of the $P$ parameter in PT05 equations, in order to interpret those high metallicity knots, we have created another model in two dimensions for the galaxy, but for the parameter $P$, employing again a Delaunay triangulation.
Form Fig. \ref{contp628}, some of the knots seem to be "caused" for a larger value in $P$. We have to take into account that for regions with the same value of $R_{23}$, a larger value of $P$ means a larger metallicity. 

These knots seem to be associated to the arms, since the mean azimuthal distance to them are of $\sim$0.4\,kpc. The parameter $P$ is directly related with the temperature of the ionizing cluster of the H~II region (see Pilyugin 2001a). For higher values of $P$, the temperature increases. So it is possible for the knots in Fig. \ref{contp628} to be due to a presence of more high temperature stars in the clusters that ionize the H~II region or, in other words, that the initial mass function (IMF) for those knots contain a larger fraction of more massive stars when compared with the rest of the H~II regions of the disk.
Variations of IMFs between galaxies, linked to the strength of the density wave, has been previously reported (Cedr\'es et al. 2005), that these kind of variations are not entirely unexpected.


\subsection{The metallicity of NGC~6946}
\begin{figure}
 \centering
   \resizebox{\hsize}{!}{\includegraphics{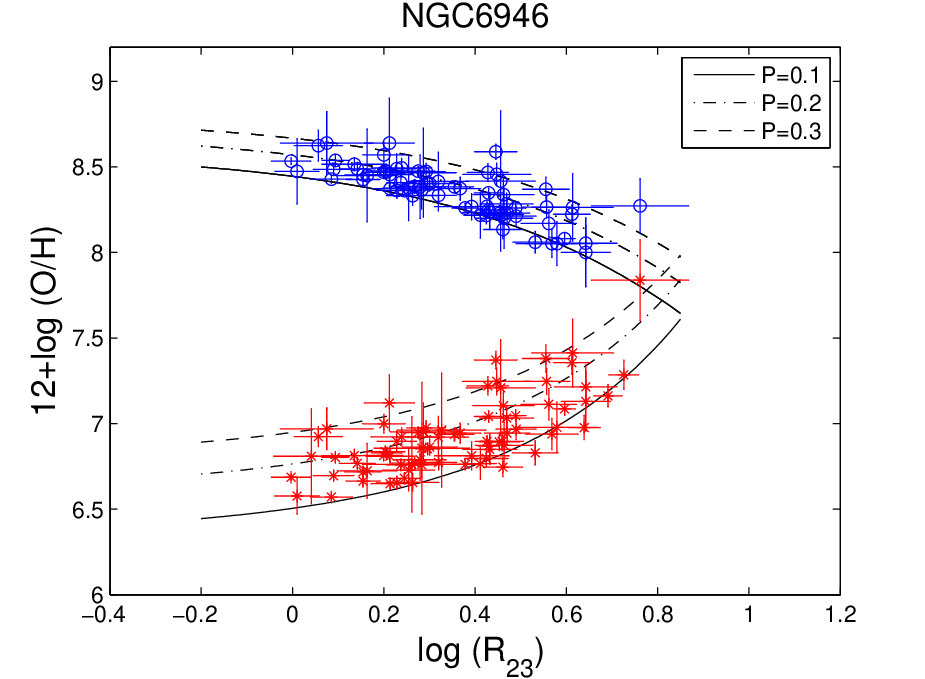}}
 \caption{Oxygen abundance for NGC~6946 as a function of the logarithm of $R_{23}$ for different values of $P$. Open circles are the regions with the metallicity derived using the upper branch calibration. Asterisks are the regions with metallicity derived using the lower branch calibration.}
 \label{metvsr236946}
\end{figure}

Unfortunately, for NGC~6946 there is not a study with the same emission lines as the present one, like the excellent work of RO11 for NGC~628. Moreover, we were not able to find in the literature data with information in the auroral oxygen lines to derive the metallicity with the direct method, so we are restricted to make comparisons only with the results from Moustakas et al. (2010).
From Fig. \ref{pvsr23} it is clear that the H~II regions of NGC~6946 have a simillar behaviour as the NGC~628 ones, with low values of $P$, so in the determination of their oxygen abundances we need only to use the upper branch of the PT05 calibration (Fig. \ref{metvsr236946}). 

\begin{figure}
 \centering
   \resizebox{\hsize}{!}{\includegraphics{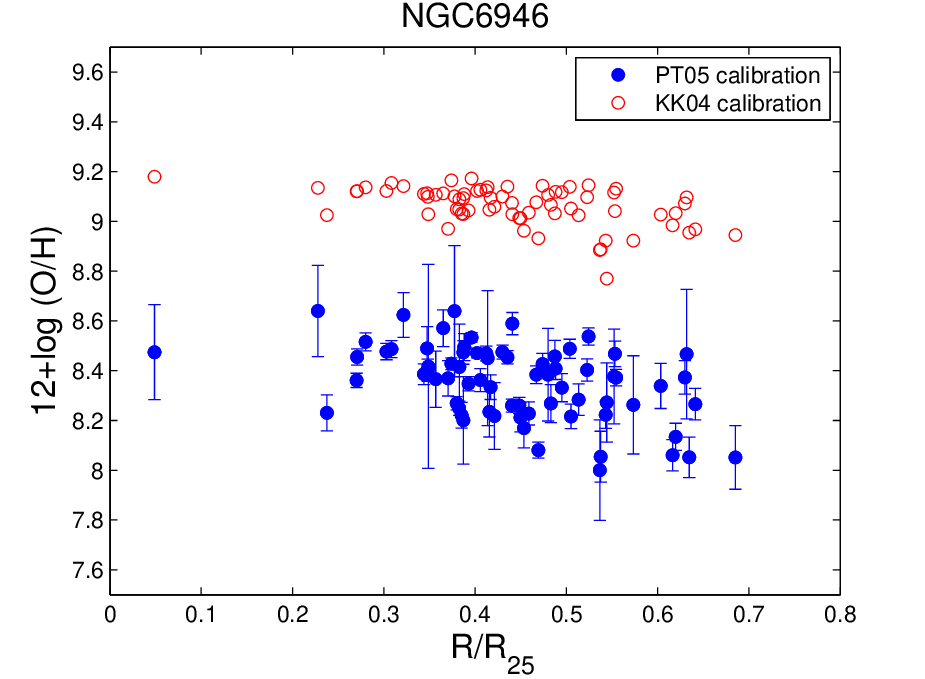}}
 \caption{Oxygen abundance for NGC~6946 as a function of the galactocentric radius, employing the calibration from PT07 (filled circles) and KK04 calibration (open circles).}
 \label{metco6946}
\end{figure}

In Fig. \ref{metco6946} we have represented the metallicity for NGC~6946, derived through PT05 calibration (filled circles) and KK04 calibration (open circles). The behaviour of the metallicity is the same as for NGC~628, with KK04 calibration yielding large values for metallicity than the ones obtained using PT05 calibration. The difference between both is about 0.5-0.6\,dex as maximun.

In table \ref{grad6946} a summary of the gradients and central oxygen abundances obtained for NGC~6946 and the results from Moustakas et al. (2010) is shown. There is a large discrepancy between our results and Moustakas et al. (2010) ones, in the gradient derived using PT05 calibration. This may be due to the small amount of H~II regions (only 6 for this galaxy) employed by Moustakas et al. (2010) in order to derive the gradient. However, taking into account the uncertainties, both results are still compatible.
Such discrepancy does not exist for KK04 calibration, probably due to the lower dispersion of the data.

\begin{table*}
\caption{Radial metallicity gradients and central oxygen abundances for NGC~6946}
\begin{tabular}{c |c c c c}
\hline\hline
 & \multicolumn{2}{c}{Gradient (dex/R$_{25}$)} & \multicolumn{2}{c}{Central metallicity}\\
 & PT05 & KK04 & PT05 & KK04\\
This work & $-$0.40$\pm0.1$ & $-$0.29$\pm0.1$ & 8.57$\pm0.04$ & 9.19$\pm0.04$\\
Moustakas et al. (2010) & $-$0.17$\pm0.15$ & $-$0.28$\pm0.1$ & 8.45$\pm0.06$ & 9.13$\pm0.04$\\
\hline\hline
\end{tabular}
\label{grad6946}
\end{table*}

We have created a two dimensional model for the oxygen abundance for NGC~6946, with the same conditions as we did for NCG~628.

\begin{figure*}[h!]
 \centering
 \begin{tabular}{c c} 
   \resizebox{9cm}{!}{\includegraphics{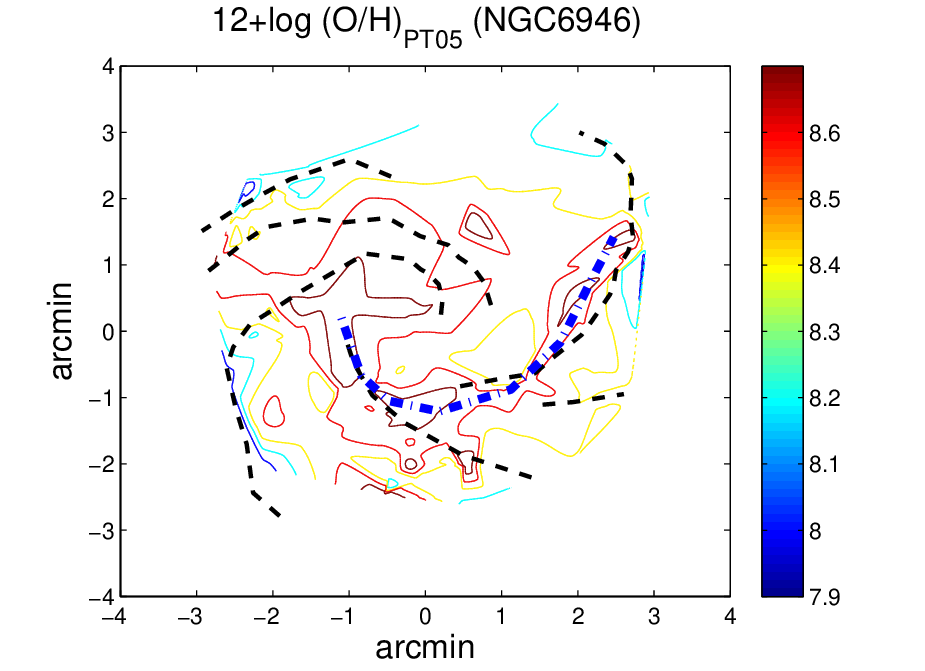}} & \resizebox{9cm}{!}{\includegraphics{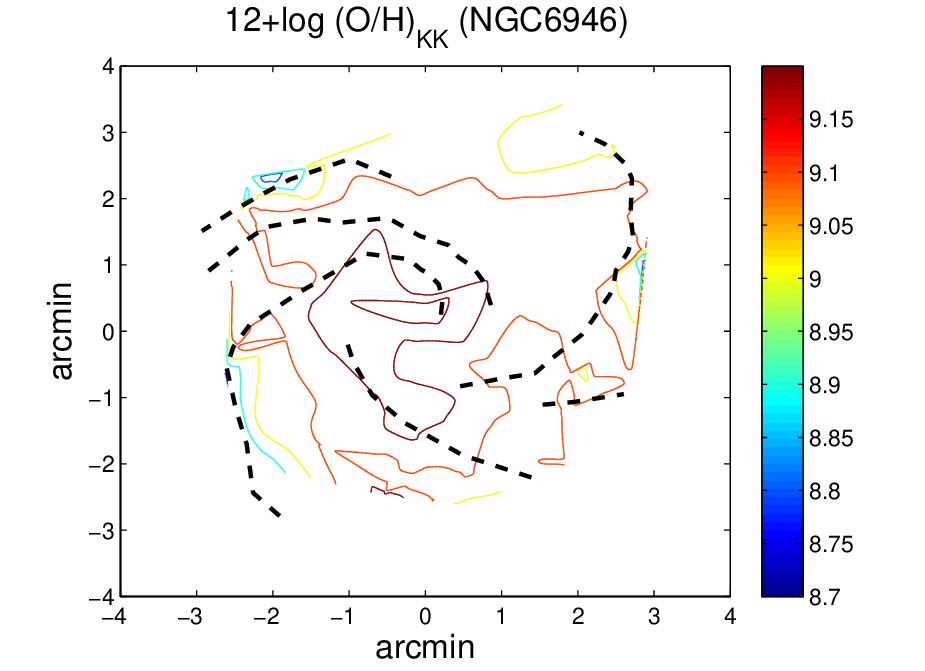}}\\
 \end{tabular}
 \caption{Contour plot of the metallicity, determined using the PT05 calibration (left panel) and the KK04 calibration (right panel), for NGC~6946. North is at the top, East is at the left. Spiral arms are indicated by the dashed lines (Cedr\'es et al. 2012). The high metallicity arm is indicated in the PT05 calibration figure by the point-dashed line.}
 \label{contmp6946}
\end{figure*}

As seen in Fig. \ref{contmp6946}, there is a clear azimuthal structure in the behaviour of the metallicity for NGC~6946. This structure is clearer in the data from the PT05 calibration, where a high metallicity spiral arm seems to exist. This higher metallicity arm seems to be associated to several stellar arms, first to the North and North-South arms, and then following the South and South-North arms. This behaviour also exists in the KK04 calibration data, although is less clear.

\begin{figure*}[h!]
 \centering
 \includegraphics{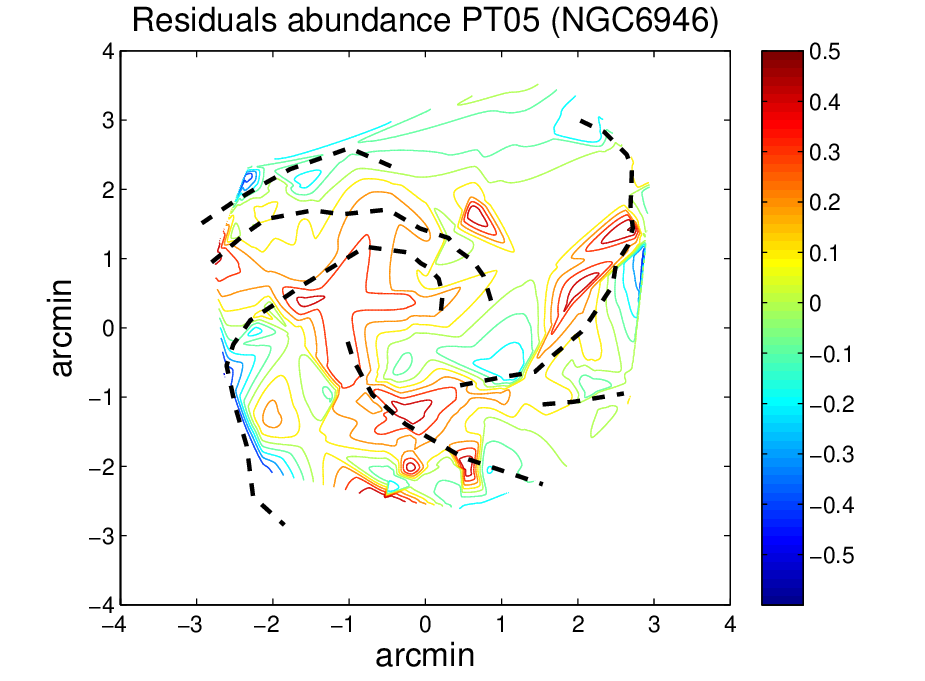}
 \caption{Contour plot of the residuals of the oxygen abundance, determined using the PT05 calibration, after the subtraction of the radial metallicity gradient, for NGC~6946. North is at the top and East is a the left. Spiral arms are indicated by the dashed lines (Cedr\'es et al. 2012).}
 \label{res6946}
\end{figure*}

In Fig. \ref{res6946} we obtain a clearer view of the azimuthal variations for NGC~6946 after subtracting the metallicity radial gradient.

\begin{figure*}[h!]
 \centering
 \includegraphics{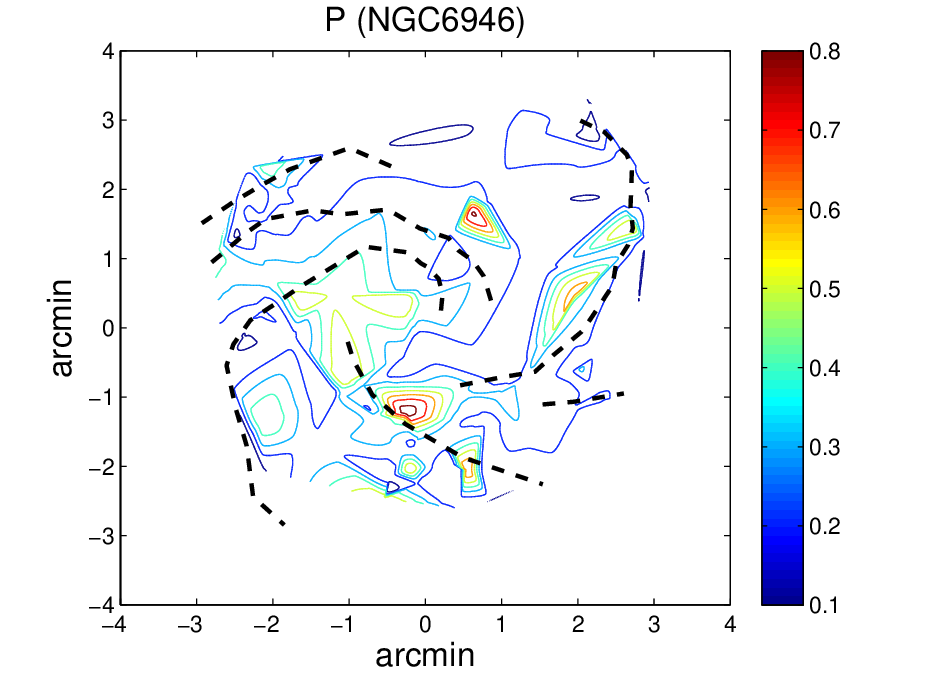}
 \caption{Controur plot of the parameter $P$ for NGC~6946. North is at the top, East is at the left. Spiral arms are indicated by the dashed lines (Cedr\'es et al. 2012).}
 \label{contp6946}
\end{figure*}

Again, in order to explore this behaviour, we have created a model for the parameter $P$, using the same conditions as the ones employed for the oxygen abundance. The results are in Fig. \ref{contp6946}. There are several knots with high $P$ that follow the same path as the higher metallicity arm, so as in the case of NGC~628 with the knots, the higher metallicity arm seems to be generated by the behaviour of $P$, that generates an arm that it is associated to a main stellar arm but that does not follow exactly their same path. As it is suggested in Cedr\'es et al. (2012) (and references cited there), there is a difference in the position between the stellar arm and the H~II regions, as it is more likely to find star forming regions "behind" the stellar arms (if we considered the direction of rotation of the galaxy) instead of being in the center of them. 

The same case as for NGC~628, can be done for this galaxy in order to explain the "arm" in $P$ and oxygen abundance: it could be due to an accumulation of regions with a slighty different IMF than those in the rest of the disk of the galaxy.

\section{Conclusions}
We have obtained data from 209 and 226 H~II regions from the galaxies NGC~628 and NGC~6946, respectively, in the emission lines: H$\alpha$, H$\beta$, [O~II]$\lambda$3727 and [O~III]$\lambda\lambda$4959,5007 and their respective continua.
For these regions the isophotal area, extinction, equivalent width and oxygen abundance were derived.

We have found that the H$\alpha$ equivalent width for NGC~6946 is slighty larger than the one derived for NGC~628. This result could be due to the presence of, on average, older H~II regions in NGC~628 when compared with NGC~6946. In general, the radial behaviour of the equivalent width for both galaxies agree with the trend observed in other galaxies (Cedr\'es \& Cepa, 2002).

The line ratios [O~II]/H$\beta$ and [O~III]/H$\beta$ were obtained for regions where the [O~II] and [O~III] emission were detectable (3\,$\sigma$ over the background noise). Our data was compared then with the results present in literature and a general agreement was found. Although we were able to obtain data for a larger number of H~II and with less investment in observing time, when compared with other works.

Radial oxygen abundance gradients for both galaxies, employing the PT05 and KK04 calibration,  were obtained. Those gradients were compared with the ones derived from the literature and again a general agreement was found.

As expected, the KK04 method yields a larger metallicity and a different gradient when compared with the PT05 method.
Even if the PT05 method presents a larger dispersion than that for the KK04 method, the gradient and the metallicity values for single regions are closer to the direct-method results, as it was previously suggested in the literature (e.g. Cedr\'es et al., 2004 or L\'opez-S\'anchez \& Esteban, 2010).

Two dimensional contour models for the metallicity (and associated parameters) for both galaxies, employing the Delaunay triangulation, were created.
Azimuthal structures for both galaxies were discovered for the metallicity: for NGC~628, there were high oxygen abundance knots linked to the arms; for NGC~6946 a high metallicity "arm" was found. The knots for NGC~628 are aproximately in the same position as those presented in S\'anchez et al. (2011).
These assymetries are not as important when the KK04 calibration is used, and appear to be linked to the behaviour of the parameter $P$.

Taking into account the relationship that exists between $P$ and the effective temperature of the ionizing clusters, it is possible that those high oxygen abundance/high $P$ regions may indicate a higher value of the effective temperature, and therefore, a higher fraction of massive stars when compared with regions with lower $P$, which also indicates a variation in the IMF for those regions.
If we take into account that the high metallicity knots (and the high metallicity arm) are associated to the galaxies' stellar arms, we propose a relationship between them and the density wave. This result support the conclusion presented in Cedr\'es et al. (2004), where it was suggested that the density wave may have an influence in the IMF. However, in order to confirm this, more observations, with galaxies with differend kinds of arms, are necessary. 


\begin{acknowledgements}
      This work was supported by the Spanish Ministry of Economy and Competitiveness (MINECO) under the grant AYA2011-29517-C03-01.
Based on observations made with the Nordic Optical Telescope, operated on the island of La Palma jointly by Denmark, Finland, Iceland, Norway, and Sweden, in the Spanish Observatorio del Roque de los Muchachos of the Instituto de Astrof\'{\i}sica de Canarias.
\end{acknowledgements}

\end{document}